\documentclass[usegraphicx, usenatbib, useAMS]{mn2e}
\usepackage{times}
\usepackage{aas_macros}

\newcommand{\muB}        {\mu_{\rm B}}

\newcommand{\RHo}        {R_{\rm Ho}}
\newcommand{\Mt}        {M_{\rm t}}
\newcommand{\Rt}        {R_{\rm t}}
\newcommand{\Rd}        {R_{\rm d}}
\newcommand{\Vc}        {V_{\rm c}}
\newcommand{\Rtw}        {R_{\rm tw}}
\newcommand{\Md}        {M_{\rm d}}

\newcommand{\omegap}        {\omega_{\rm p}}
\newcommand{\omegapd}        {\omega_{\rm p,\;disc}}
\newcommand{\omegapt}        {\omega_{\rm p,\;torus}}

\newcommand{\Ptwo}        {P_2(\cos \theta)}
\newcommand{\Pfour}        {P_4(\cos \theta)}
\newcommand{\GMoR}        {\frac{G\Mt}{\Rt}}
\newcommand{\roRt}        {\frac{r}{\Rt}}
\newcommand{\tauave}        {\langle\tau\rangle}

\newcommand{\dphi}         {\Delta \Phi_{\rm p}}
\newcommand{\dphih}        {\frac{\Delta \Phi_{\rm p}}{2}}
\newcommand{\philon}       {\phi_{\rm LON}}
\newcommand{\Philon}       {\Phi_{\rm LON}}
\newcommand{\sini}         {\sin i \;}
\newcommand{\cosi}         {\cos i \;}
\newcommand{\sinis}        {\sin^2 i \;}

\newcommand{\sinht}        {\sin\frac{\Theta}{2}}
\newcommand{\sinhp}        {\sin \dphih}

\newcommand{\sinhps}       {\sin^2 \dphih}

\newcommand{\AB}           {\overline{AB}}

\newcommand{\ie}{\textrm{i.e.}}
\newcommand{\eg}{\textrm{e.g.}}

\begin{document}
\title[Warps induced by cosmic infall]{Galactic warps induced by
cosmic infall}

\author[Juntai Shen and J. A. Sellwood]{Juntai Shen\thanks{Currently Harlan
Smith Fellow, McDonald Observatory, The University of Texas at Austin,
1 University Station C1402, Austin, TX 78712; email:
shen@astro.as.utexas.edu} and J. A. Sellwood\thanks{Email:
sellwood@physics.rutgers.edu}\\ Department of Physics and Astronomy,
Rutgers University, 136 Frelinghuysen Road, Piscataway, NJ 08854, USA}

\date{Accepted \today}
                                                                               
\pagerange{\pageref{firstpage}--\pageref{lastpage}} \pubyear{2006}
                                                                                
\maketitle
                                                                               
\label{firstpage}

\begin{abstract}
Recent ideas for the origin and persistence of the warps commonly
observed in disc galaxies have focused on cosmic infall.  We present
$N$-body simulations of an idealized form of cosmic infall onto a disc
galaxy and obtain a warp that closely resembles those observed.  The
inner disc tilts remarkably rigidly, indicating strong cohesion due to
self-gravity.  The line of nodes of the warp inside $R_{26.5} \sim 4.5
\Rd$ is straight, while that beyond $R_{26.5}$ generally forms a
loosely-wound, leading spiral in agreement with Briggs's rules.  We
focus on the mechanism of the warp and show that the leading spiral
arises from the torques from the misaligned inner disc and its
associated inner oblate halo. The fact that the line of nodes of most
warps forms a leading spiral might imply that the disc mass is
significant in the centre. If the line of nodes can be traced to very
large radii in future observations, it may reveal information on the
mass distribution of the outer halo. The warp is not strongly damped
by the halo because the precession rate of the inner disc is slow and
the inner halo generally remains aligned with the inner disc.  Thus
even after the imposed quadrupolar perturbation is removed, the warp
persists for a few Gyrs, by which time another infall event can be
expected.
\end{abstract}


\section{Introduction}
\label{sec:intro}
The optically visible parts of galactic discs are usually remarkably
thin and flat, whereas the more extended HI discs of many edge-on
galaxies appear noticeably warped with an integral sign shape.
Stellar warps do exist in some galaxies \citep[e.g.][]{res_etal_02},
but are much less pronounced than the warps in the extended gaseous
discs.  We infer that warps are a gravitational phenomenon because
weak stellar warps, if present, follow the same warped plane as the
gaseous ones \citep{cox_etal_96}.

Warps are extremely common -- at least half of all spiral galaxies are
warped: \citet{bosma_91} found 12 out of 20 edge-on spiral galaxies
are warped.  The more recent HI observations by \citet{gar_etal_02}
found that 20 out of 26 edge-on galaxies are warped.  Since these
surveys detected only warps whose line of nodes is close to our line
of sight, the true fraction of warps must be even higher. In fact
\citet{gar_etal_02} also found that all galaxies possessing an HI disc
more extended than the optical one are warped. The ubiquity of warps
suggests that warps are either repeatedly regenerated or long-lived.

Warps can be detected kinematically even when the system is not
edge-on.  \citet{briggs_90} studied a sample of 12 warped galaxies
with high-quality 21-cm data, and found that galactic warps obey some
fairly simple rules:

\begin{enumerate}

\item the HI layer typically is coplanar inside radius $R_{25}$, the
radius where the B-band surface brightness is $25\,\hbox{mag
arcsec}^{-2}=25\muB$.\footnote{If the disc is exponential with
scale-length $\Rd$ and central surface brightness $I_{B,0} \sim 21.7
\, {\rm mag \, arcsec}^{-2} = 21.7 \muB$ \citep[the so-called
Freeman's law]{freema_70}, $R_{\rm 25} \sim 3.0 \Rd$ and $R_{\rm 26.5}
\sim 4.4 \Rd$.} and the warp develops between $R_{25}$ and $R_{26.5}$
(aka the Holmberg radius);

\item the line of nodes (LON) is roughly straight inside $R_{\rm 26.5}$;

\item the LON takes the form of a loosely-wound {\em leading spiral}
outside $R_{\rm 26.5}$.

\end{enumerate}

The origin and persistence of warps still presents a puzzle.
\citet{hun_too_69}, before the discovery of dark matter halos, studied
the bending dynamics of isolated thin discs.  They found that discrete
warp modes in a cold, razor-thin disc do not exist unless the edge is
unrealistically sharp.  Such modes (standing waves) can be realized by
the superposition of outgoing and ingoing waves, provided that waves
incident on the edge can reflect.  However, a disc with a smooth edge
does not reflect bending waves \citep{toomre_83}.

Subsequent ideas of warp formation rely in some way on the interaction
between the disc and its dark matter halo.  \citet{dek_shl_83} and
\citet{toomre_83} suggested that a flattened halo misaligned with the
disc could form a long-lasting warp.  \citet{spa_cas_88} and
\citet{kuijke_91} invoked a rigid flattened halo and obtained
persisting warps (dubbed ``modified tilt modes''), which were
insensitive to the details of the disc edge.  \citet{lovela_98}
studied the tilting dynamics of a set of rings also in a rigid halo,
but generally assumed that the inner disc lay in the symmetry plane of
the spheroidal halo.

Dark matter halos are not rigid, however, and a responsive halo alters
the dynamics in several ways.  \citet{nel_tre_95} showed that the
precession of a misaligned disc inside a flattened halo would sap
energy from the warp mode through dynamical friction, damping it on
timescales much shorter than a Hubble time.  Numerical studies using
$N$-body halos also found the discrete warp mode does not survive:
\citet{dub_kui_95} found that the inner halo and disc quickly settle
into alignment.  \citet{bin_etal_98} confirm that a discrete warp mode
in a rigid halo does not survive in a simulation with a responsive
halo, and conclude that the inner halo could never be significantly
misaligned from the inner disc.

\citet{ost_bin_89} drew attention to the likely misalignment with the
disc axes of late infalling material in hierarchical galaxy formation
models, and proposed that warps arise due to the slewing of the
galactic potential as material with misaligned angular momentum is
accreted.  \citet{qui_bin_92} showed, for a realistic cosmic scenario,
that the mean spin axis of a galaxy must slew as late arriving
material rains down on the early disc.  The less-than-critical matter
density in modern $\Lambda$CDM universe models implies that infall is
less pervasive at later times, but it manifestly continues to the
present day in gravitationally bound environments such as that of the
Local Group.

\citet[hereafter JB99]{jia_bin_99} present results of an experiment
in which a disc is subjected to the torque from a misaligned, massive
torus at a large radius.  This well-defined perturbation is a very
crude model of an outer halo that is rotationally flattened, with a
spin axis misaligned with that of the disc.  It is misaligned and
farther out because, in hierarchical scenarios, the mean angular
momentum of the later arriving outer halo is probably misaligned from
that of the original inner halo and disc.  They concede that the
accretion axis is in reality likely to slew continuously over time, so
a model with a constant inclination is somewhat unrealistic.

Here we use high quality $N$-body simulations of a model of this type.
We improve on the work by JB99 in the following main aspects:

\begin{itemize}
\item We use a disc of particles with random motion, whereas JB99
employed a disc composed of rigid rings coupled only by gravitation.
Thus JB99 did not include random motion of the disc stars, which has
been shown to add to the stiffness of the disc \citep{deb_sel_99};

\item We use a much more extensive disc so that we can study the warp
-- JB99 truncated the disc at $4\Rd$.  The behaviour of the LON beyond
this radius is important for comparing our results with observations
of warps (\eg\  Briggs's rules);

\item In \S\ref{sec:warppersist} we remove the forcing perturbation at
later times and follow the evolution of the distorted disc fully
self-consistently.

\end{itemize}

\noindent We obtain flat inner discs and long-lasting warps in the
outer disc that match all of Briggs's rules quite well.  We present a
detailed analysis of the dynamics and show, for example, that the
persistence of warps is not nearly as perplexing as is currently
believed.

\section{Model setup and simulation details}
\label{sec:modelsetup}
\begin{figure} 
\vspace{0.02\vsize}
\centerline{\includegraphics[angle=0,
width=.9\hsize]{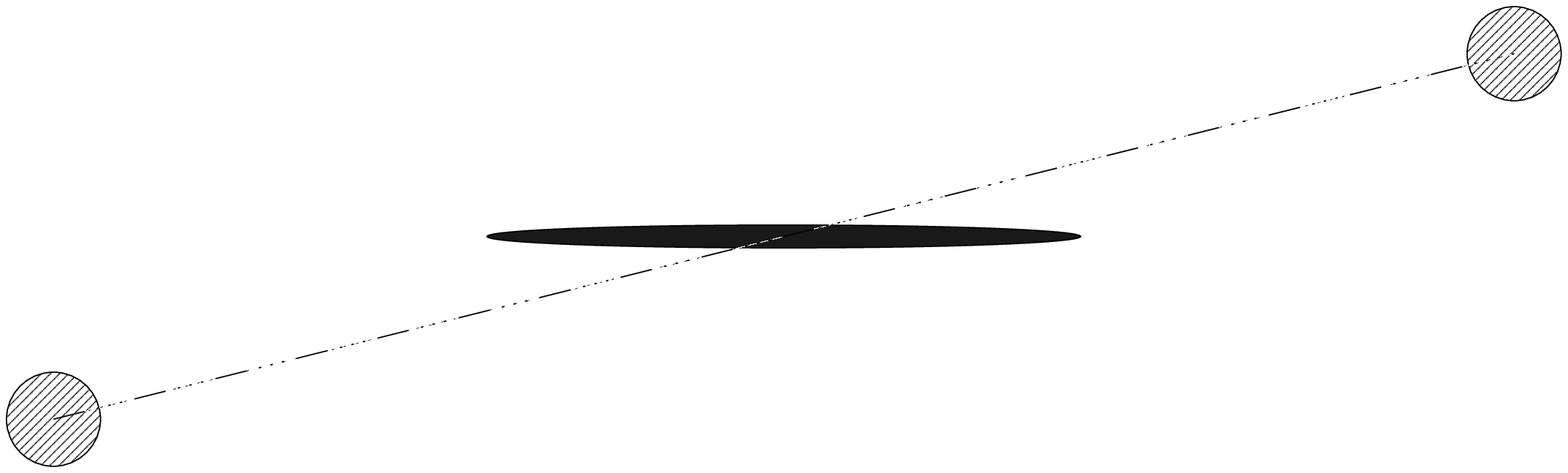}}
\caption[A sketch of initial model setup.]
{A sketch of initial model setup.  The dark matter halo is not shown.}
\label{fig:initial}
\vspace{0.02\vsize}
\end{figure}

Following JB99, our simulations include three distinct mass
components: a disc, a halo, and an accreted torus.  The plane of the
torus is inclined to that of the disc, as sketched in
Figure~\ref{fig:initial}.

Our simulations are set up as follows: The initial disc has the
exponential surface density profile:
\begin{equation}
\Sigma(R)= \frac{\Md}{2\pi \Rd^2}  \exp{\left( - \frac{R}{\Rd}\right)},
\label{eqn:expdisk}
\end{equation}
where $R$ is the cylindrical radius and $\Rd$ and $\Md$ are the scale
radius and the total mass of the disc, respectively.  We truncate the
disc at $R=8\Rd$, spread the particles vertically as the isothermal
sheet with a locally-defined equilibrium vertical velocity dispersion,
and embed it in the halo; the rms vertical thickness of the initial
disc is $0.1\Rd$ at any radius.  We set the initial velocities of the
disc particles such that the Toomre stability parameter $Q\simeq 1.5$
and the disc is in rotational balance.

The halo is set up so as to be in equilibrium with the disc using the
procedure described in Appendix A of \citet{deb_sel_00}.  The
distribution function has a King-model form with $\Psi(0)/\sigma^2 =
2.0$, where $\sigma$ is the central velocity dispersion and $\Psi(0)$
is the relative potential at the centre \citep{bin_tre_87}, which in
our case includes a contribution from the disc.  We set the tidal
radius $r_t=18\Rd$, and total mass $= 9.0 \Md$.  The King radius is
$r_{c}\sim 5.6 \Rd$ and the half-mass radius of the halo is $r_h \sim 5
\Rd$.  The inner halo is mildly oblate initially due to the gravity of
the disc. The halo we adopted here is massive, as we attempt to use a
similar halo profile used by JB99 in which $M_{\rm h}/M_{\rm d}=5.0$
for $r<3.2\Rd$.  Other experiments with halos having different mass
profiles, to be reported elsewhere, have revealed that the conclusions
presented here are not strongly affected by the choice of halo;
warping behaviour is less sensitive to the density profile or total
mass of the halo than to the masses and sizes of the disc and torus.

Following JB99, we gradually inject particles into an inclined torus,
causing its mass to increase linearly from 0 at $t=0$, to $\Mt=2.5\Md$
at $t=200$.  The torus, which has radius $\Rt$, is uniform in azimuth
and has Gaussian density profile in cross-section, with a standard
deviation $b$; in our case, $\Rt=15.0\Rd$ and $b=0.2\Rd$.  The torus
plane is inclined at $15^\circ$ relative to the initial disc plane and
we adopt the line of nodes as the $y$-axis of our coordinate system.

The initial velocities of injected torus particles are equal to the
local circular speed in magnitude and directed tangentially around the
torus, as in JB99.  Thus the torus has very large angular momentum,
and therefore does not precess significantly from its original plane
in response to the torque from the disc.  However, a cold torus is
dynamically unstable and breaks into radially extensive spiral
filaments after a few Gyrs when allowed to evolve self-consistently.
The breakup of the cold torus depends on many factors that we do not
fully understand --- e.g., tori with the same parameters dissolve
differently and at different times in runs with rigid and live halos.
In order to avoid the inconvenient complication of time-varying,
non-axisymmetric perturbing forces acting on the disc, we generally
keep the torus particles fixed in their initial positions relative to
the galactic centre.  This stratagem enables us to compare the
dynamical evolution of warps in different experiments with the same
torus parameters.

All 1.25 million particles belonging to massive components in our
simulations have equal masses.

We adopt $\Rd$ and $\Md$ as our units of length and mass,
respectively, and our time units are therefore dynamical times
$(\Rd^3/G\Md)^{1/2}$.  Generally all quantities are expressed in units
such that $G=\Md=\Rd=1$ unless otherwise noted.  These units can be
scaled to physical values as desired; we adopt one possible scaling,
choosing $\Rd = 2.5\;$kpc and a unit of time of $10\;$Myr implies $\Md
= 3.47 \times 10^{10}M_\odot$; our unit of velocity scales to
244~km~s$^{-1}$.

We use the hybrid particle-mesh scheme described in detail in
\citet[Appendix~B]{sellwo_03}.  The self-gravity of the disc is
computed on a high-resolution cylindrical polar grid
\citep[see][]{sel_val_97, she_sel_04}, while that of the halo is
computed using a surface harmonic expansion on a spherical grid.  We
have carefully checked that the preferred plane of the cylindrical
polar grid does not introduce artificial forces between a tilted disc
and the grid, which requires a fine polar grid.  Our adopted numerical
parameters are summarized in Table~\ref{tab:params}.

\begin{table}
\caption{Numerical parameters used in the canonical simulation}
\begin{center}
\begin{tabular}{lrr}
\hline\hline\\
                   & Cylindrical grid       & Spherical grid \\
\hline
Grid size\dotfill        & $(N_R,N_\phi,N_z)\quad$ & $N_r = 401$\\
                   &  $ = (102,128,125)$      \\
Angular compnts\dotfill   & $0\leq m \leq 8$       & $0 \leq l \leq 4$ \\
Outer radius ($\Rd$)\dotfill      & 8.0               & 18.0 \\
$z$-spacing ($\Rd$)\dotfill       & 0.02 \\
Softening length ($\Rd$)\dotfill  & 0.02  \\
$N$\dotfill              & 0.1M                     & 0.9M \\
Time step\dotfill          & 0.04                & 0.04  \\
\hline
\end{tabular}
\end{center}
\label{tab:params}
\end{table}

Over a period of 400 dynamical times, energy is conserved to about
0.02\%, angular momentum components ($L_x$, $L_y$, $L_z$) change no
more than 1\%, and the absolute change of linear momenta
($p_x$, $p_y$, $p_z$) are less than 0.06 in units we adopted.  Without
the misaligned torus, these global integrals are conserved with a
precision almost one order of magnitude better.

Our main warp diagnostic is the tip-LON diagram introduced by
\citet{briggs_90}.  We construct it from an analysis of the disc
particles, which are binned into annuli.  The inner disc is very stiff
and remains very closely coplanar within the innermost $3\Rd$; we
therefore use a single bin for all particles inside this radius and
adopt cylindrical bins (spherical bins are preferred if run for a very
long time) of equal radial widths $0.7\Rd$ outside this radius to
$R=7.9\Rd$.  (The bins remain aligned with the original disc axes.)
We then compute the eigenvectors of the inertia tensor to determine
the orientation of the disc element, $\theta$ and $\philon$ in each
bin.  We use the orientation of the inner bin to define the plane of
the inner disc.  We have experimented with various bin widths and
found this binning scheme gives us the best compromise between spatial
resolution and noise in estimation of the disc orientation (especially
for thicker/``hotter'' inner discs).  We describe the diagram and its
meaning in \S\ref{sec:leadinglon}.

\begin{figure} 
\centerline{\includegraphics[angle=0, width=.91\hsize]{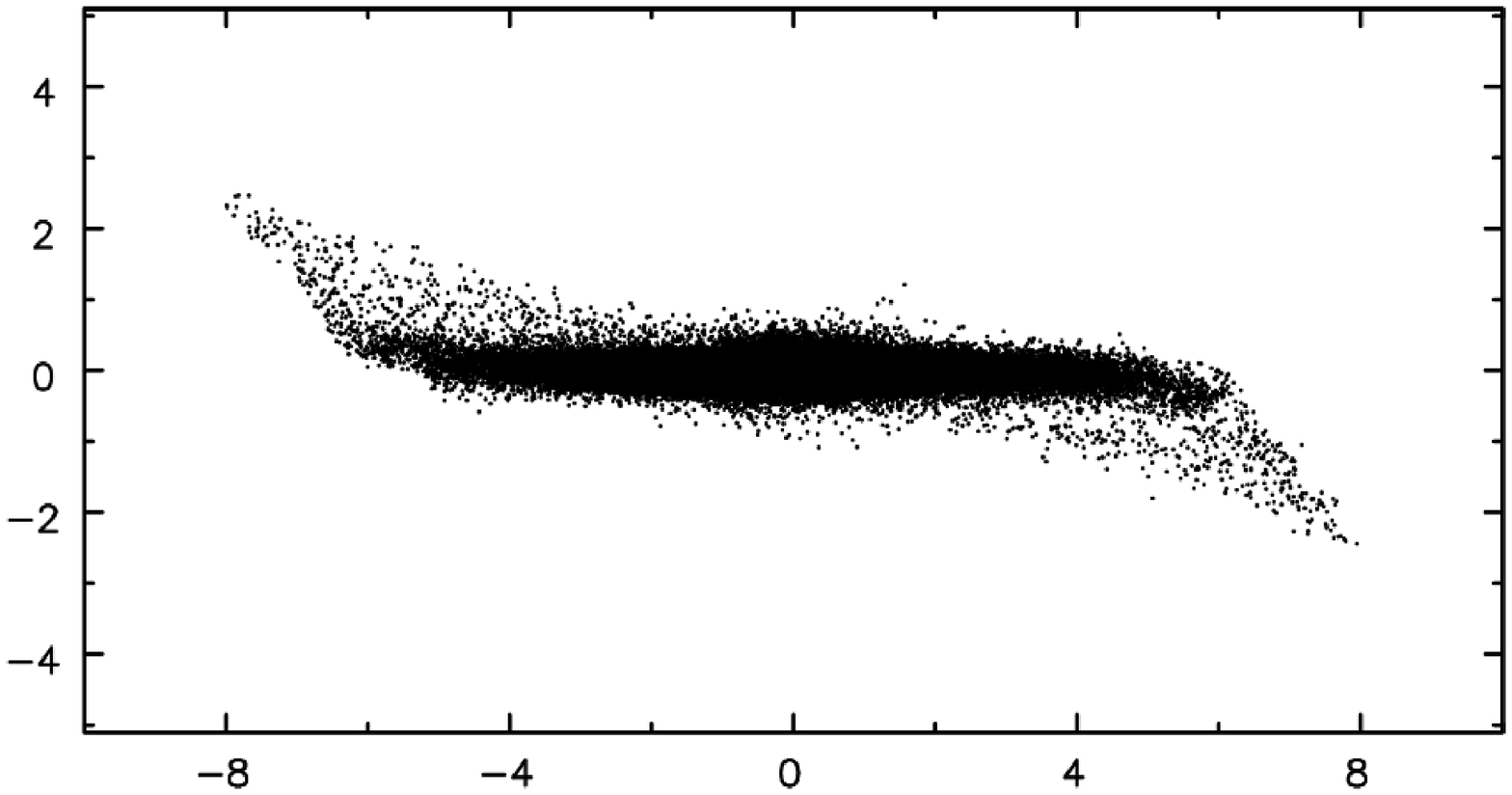}}
\centerline{\hspace{.05\hsize}\includegraphics[width=.87\hsize]{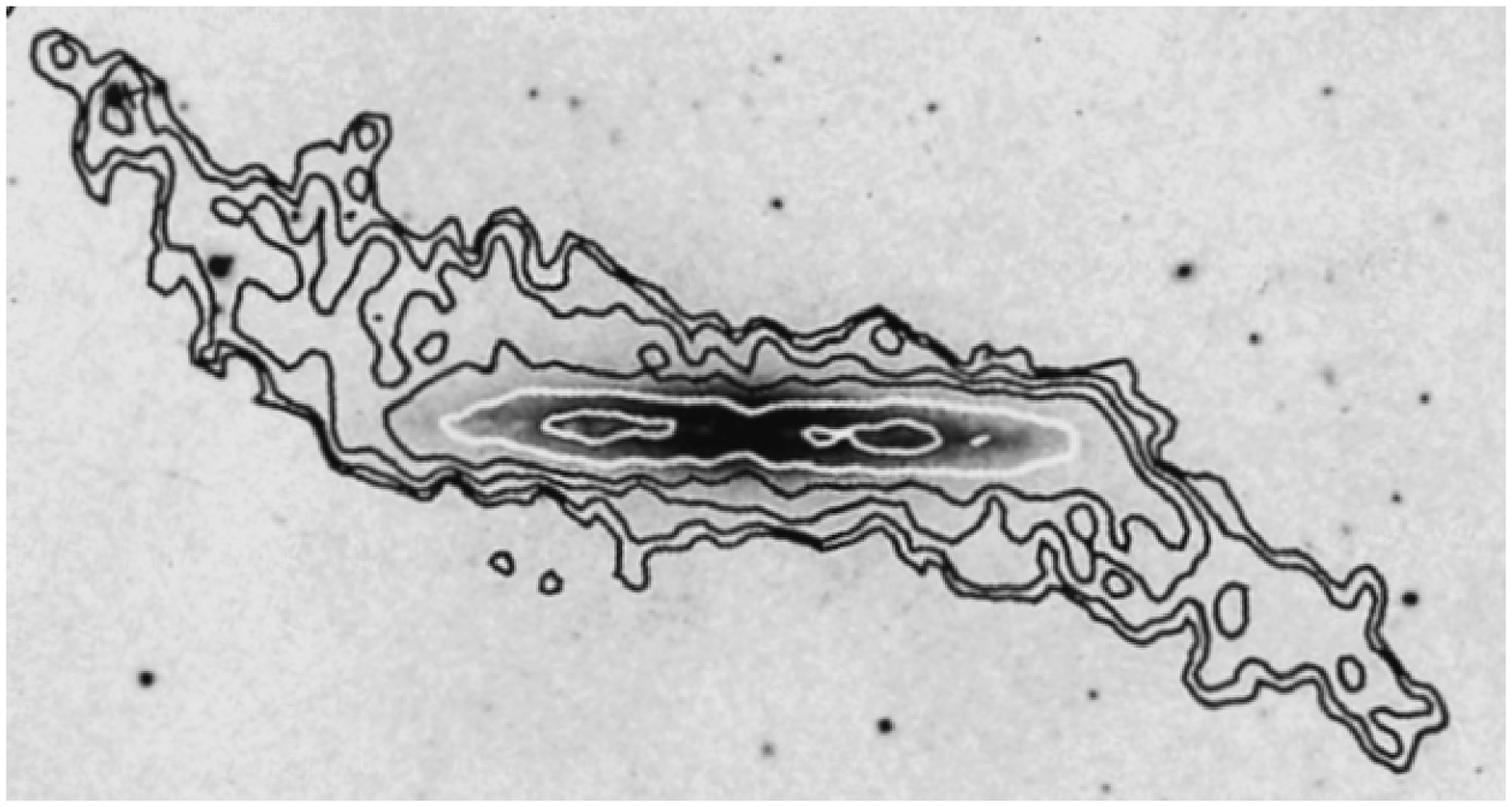}}
\caption[(a) The warp at $t=400$ in our simulation; its morphology
closely resembles the observed HI warp of NGC 4013 \citep{bottem_96},
shown in (b).]  {(a) The warp at $t=400$ in our simulation; its
morphology closely resembles the observed HI warp of NGC 4013
\citep[reproduced with permission]{bottem_96}, shown in (b). The
length unit shown is the scale length $\Rd$ of the exponential
disc. Note that we have oriented the model so that the inner
($R<3\Rd$) disc lies in the $x$-$y$ plane, which is perpendicular to
the paper.}
\label{fig:morph}
\end{figure}

\begin{figure} 
\centerline{\includegraphics[angle=-90, width=.9\hsize]{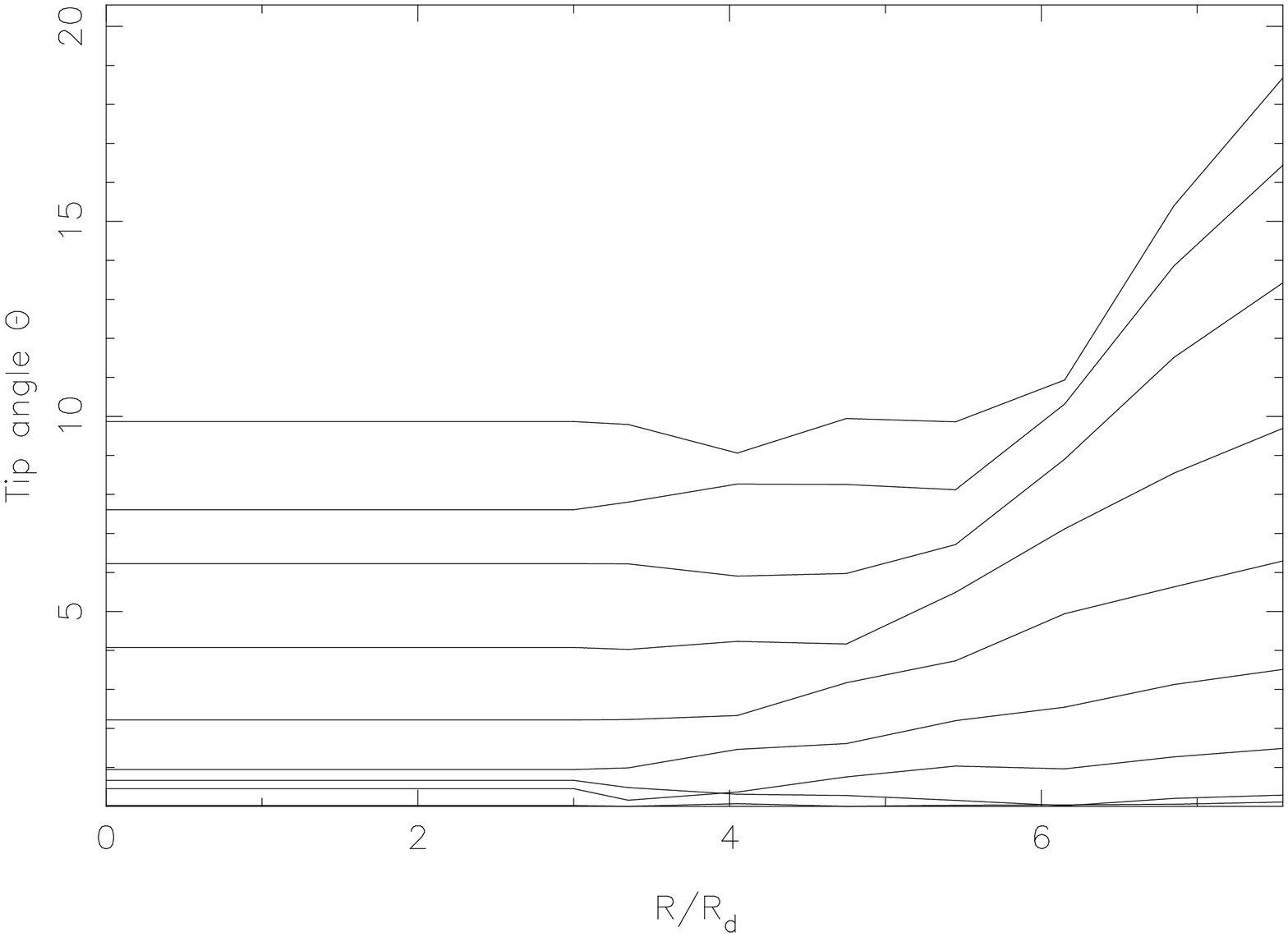}}
\vspace{.02\hsize}
\centerline{\includegraphics[angle=-90, width=.9\hsize]{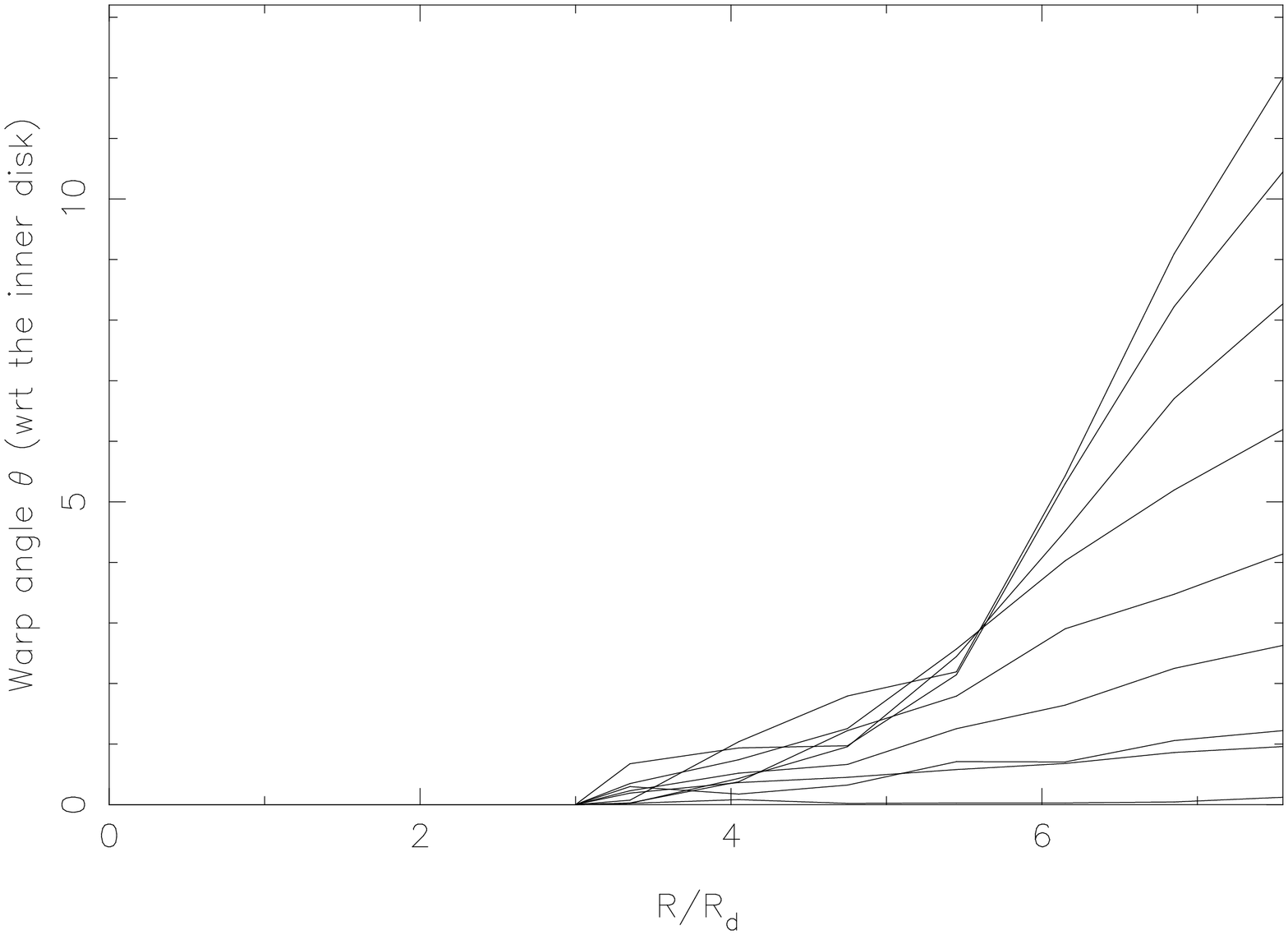}}
\caption[(a) Disc tip angle $\Theta$ as a function of radius at
intervals of 40 time units ($t=0$ to $t=320$ from bottom up). (b) as
for (a), but $\theta$ is relative to the inner disc.]
{(a) Disc tip angle $\Theta$ as a function of radius at intervals of
40 time units ($t=0$ to $t=320$ from bottom up). (b) as for (a), but
the warp angle $\theta$ relative to the inner disc.  The inner disc is
defined by the radial range $0<R<3\Rd$.}
\label{fig:thetaR}
\end{figure}

\begin{figure} 
\centerline{\includegraphics[angle=-90, width=.9\hsize]{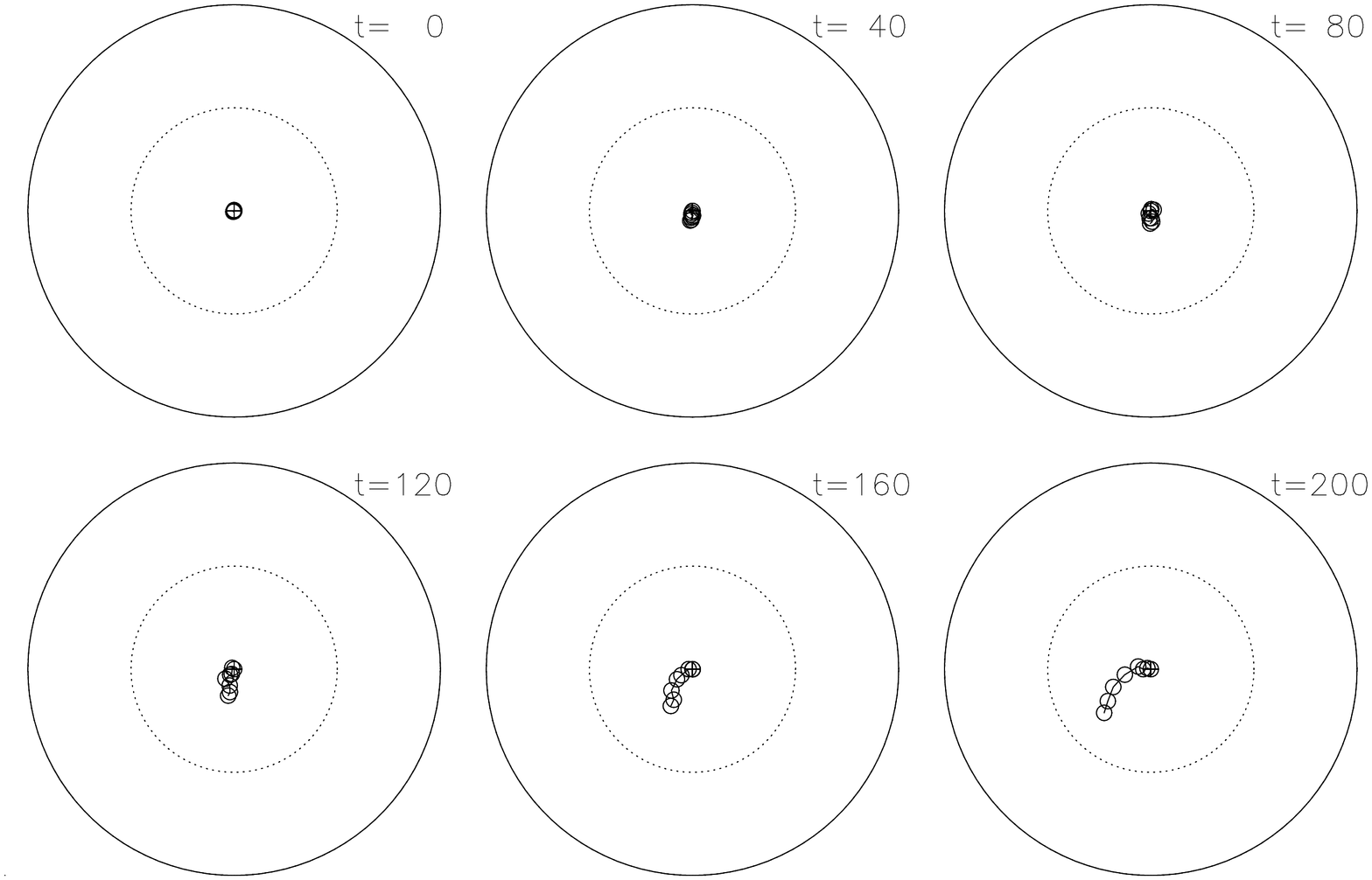}}
\vspace{.02\hsize}
\centerline{\includegraphics[angle=-90, width=.9\hsize]{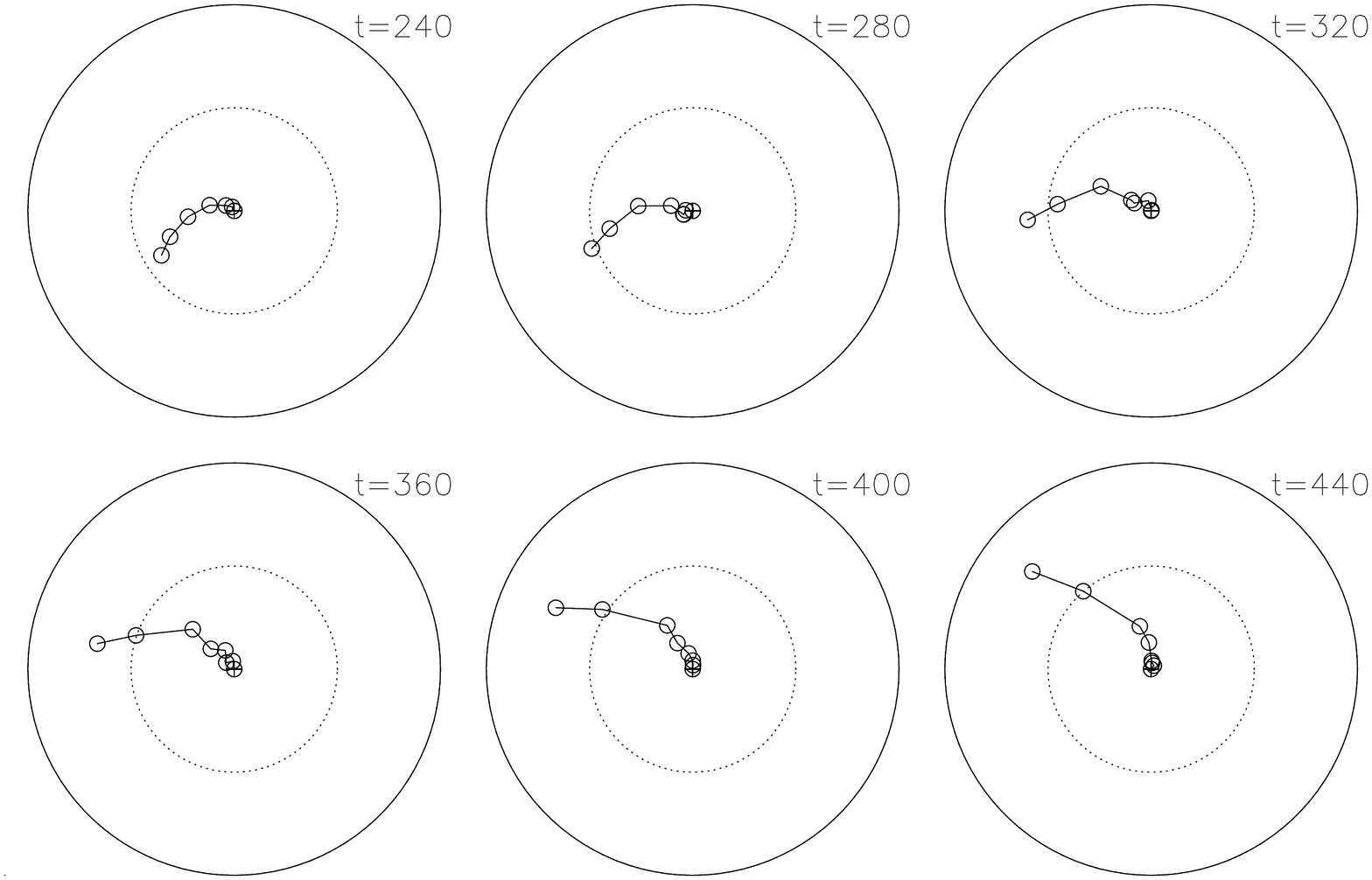}}
\caption[Tip-LON diagrams showing the time evolution of the warp in
our canonical experiment.  The LON of the warp almost always form a
leading spiral in our study, in the reference plane defined by the
inner disc.]
{Tip-LON diagrams showing the time evolution of the warp in our
canonical experiment.  The various points indicate $\theta$ (radial
coordinate) and $\philon$ (azimuthal coordinate) averaged over annular
bins having widths of $0.7 \Rd$.  The warp angle, $\theta$, is larger
for annuli with larger radii; the outermost point is for the annulus
centered on $R=7.55\Rd$.  The large circles mark warp angles of
$10^\circ$ (dotted) and $20^\circ$ (solid).}
\label{fig:typicallon}
\end{figure}

\section{A canonical simulation}
\label{sec:results}
Without the torus, the disc in our model heats mildly; $Q$ rises from
$\sim 1.5$ to $\sim1.8$ over 300 dynamical times, the disc thickens by
$\sim 20\%$ and, of course, shows no signs of warping.  The virial
ratio stays at 0.5, confirming the equilibrium of our initial model.
The disc remains unbarred until very late times.

We grow the inclined torus in our canonical experiment, in which the
initially thin and flat disc experiences a growing torque.  Since the
disc is spinning rapidly, it tends to precess about the symmetry axis
of the distant torus; differential precession will make the disc warp.
We present a quantitative discussion of the precession in
\S\ref{sec:lonanalysis}.  The much larger moment of inertia of the
accreted torus implies that its precession in response to the torque
from the disc is minimal, and we therefore keep the torus particles
frozen in their initial positions relative to the galaxy centre in
order to provide steady forcing.  Some differences in the disc
behaviour between this case and one in which the torus particles move
begin to appear after $t \sim 320$.

\subsection{Morphology of warps}
Figure~\ref{fig:morph} shows the edge-on projection of the warped disc
in the canonical simulation at $t=400$, or $200$ dynamical times after
torus growth ended.  To reveal the warp more clearly, we have rotated
the particle coordinates to the frame of the inner disc.

To illustrate that the warp in our simulation indeed resembles those
observed, we compare it to the strongly warped hydrogen layer of NGC
4013 \citep{bottem_96}.  The correspondence is not perfect; \eg, our
simulated warp starts at around $5\Rd$ whereas the warp of NGC 4013
starts to develop at the edge of its optical disc, about $4\Rd$
\citep{bottem_96}.  We could adjust the density profile of the halo,
which affects the radius at which the disc starts to warp, but have
not attempted to tailor our model to improve the match.

Note also that the particles in our simulation are collisionless
whereas the warp is observed in the dissipative gas component.
Nevertheless, the comparison should be reasonable because the
precession rate of the simulated warp is low -- collisional
dissipation in a gas layer becomes important only when the
differential precession rate is high \citep{gunn_79, toh_etal_82}.
Furthermore gas often condenses into very compact clouds, which behave
quasi-ballistically in a similar manner to a stellar system
\citep{binney_92}.

\subsection{Warp angle}
Figure~\ref{fig:thetaR}(a) shows $\Theta(R)$, the tip angle
relative to the $z$-axis of the initial disc, at various times.  In
Figure~\ref{fig:thetaR}(b) we transform to warp angles $\theta$, which
are measured relative to the orientation of inner disc ($R<3\Rd$) at
each time.  The warp amplitude at later times in
Figure~\ref{fig:thetaR}(b) seems to compare well with that in NGC
4013, both visually and quantitatively \citep[cf.][Figure
6]{bottem_96}.

\subsection{LON of the warp is a leading spiral}
\label{sec:leadinglon}
Consistent with our convention for warp angles, we use $\philon$ for
the azimuth of the line of nodes between the plane of the inner disc
and that of the annular bin, and $\Philon$ for the corresponding line
of nodes with the initial unperturbed disc plane.

Figure~\ref{fig:typicallon} shows the evolution of LON of the warp, in
the form of tip-LON plots \citep{briggs_90}.  The points (small open
circles) indicate the angle ($\theta$, $\philon$) of the symmetry axis
of an annulus of disc material, relative to the inner disc plane at
any given time.  The radial coordinate is the warp angle
(inclination), $\theta$, and the azimuthal coordinate indicates the
azimuth, $\philon$, of the best-fit plane to a disc annulus.  Points
at successively larger ring radii are joined sequentially by straight
line segments.  Note that since the warp angle, $\theta$, increases
with radius (Figure~\ref{fig:thetaR}), the plotted points refer to
radially ordered annuli from the centre out.

Figure~\ref{fig:typicallon} shows that the LON of the warp precesses
in the retrograde direction, while curving gradually into a {\em
leading} spiral (\ie, the LON advances in the direction of galaxy
rotation for successively larger radii), consistent with the third
rule of \citet{briggs_90}.  We have run many other experiments with
various parameters and found that the LON to $8\Rd$ always forms a
leading spiral when the disc is perturbed by an outer inclined torus.
The reason for the leading spiral is explained in detail in the next
section (\S\ref{sec:lonanalysis}).  Also the LON is fairly loosely
wound throughout our simulation.  \citet{bottem_96} concluded that the
LON of NGC 4013 winds modestly, by about $20^\circ$, which is
consistent with that in our simulated warp, in addition to the nice
visual resemblance in Figure~\ref{fig:morph}.

\section{In-depth analysis on the shape of LON}
\label{sec:lonanalysis}
Since there are many factors that affect the warp morphology, we break
the dynamics into parts.  We first study the simplest aspect
analytically, and then gradually add in separate pieces of the physics
to illustrate their effects.  These simplified experiments are very
helpful for developing understanding of warp dynamics and the reason
for the curved LON.  We stress that our adopted massive torus at a
very large radius has so much angular momentum that its precession due
to the torques from the disc and flattened inner halo can be
neglected.

\subsection{A test particle disc in a rigid halo}
\label{sec:toymodel}
We first study a very simple model, in which the halo is both rigid
and spherical.  We also replace the self-gravitating disc with
massless test particles and represent only the torus with massive
particles.  This model is sufficiently simple that the behaviour can
be predicted analytically and provides a welcome check of our
numerical code.  We expand the potential of the distant torus as a
series of multipoles and calculate the disc precession rate
analytically, both to second- and to fourth-order, in
Appendix~\ref{app:pre}; we give exact formulae for $\Theta$ and
$\Philon$ in Appendix~\ref{app:thetaphi}.

We use $20\,000$ test particles to represent the massless disc and
adopt a rigid, spherical King-model halo with $\Psi(0)/\sigma^2 =
6.0$, $r_t \simeq 18.0$, and total mass $9.0$.  Only the torus is
composed of massive particles, and is grown from $\Mt=0$ to $2.5$
within $200$ time units, as before.  The inclination angle of the disc
with respect to the plane of the accreting torus is $i = 15^\circ$.

\begin{figure} 
\centerline{\includegraphics[angle=0, width=.9\hsize]{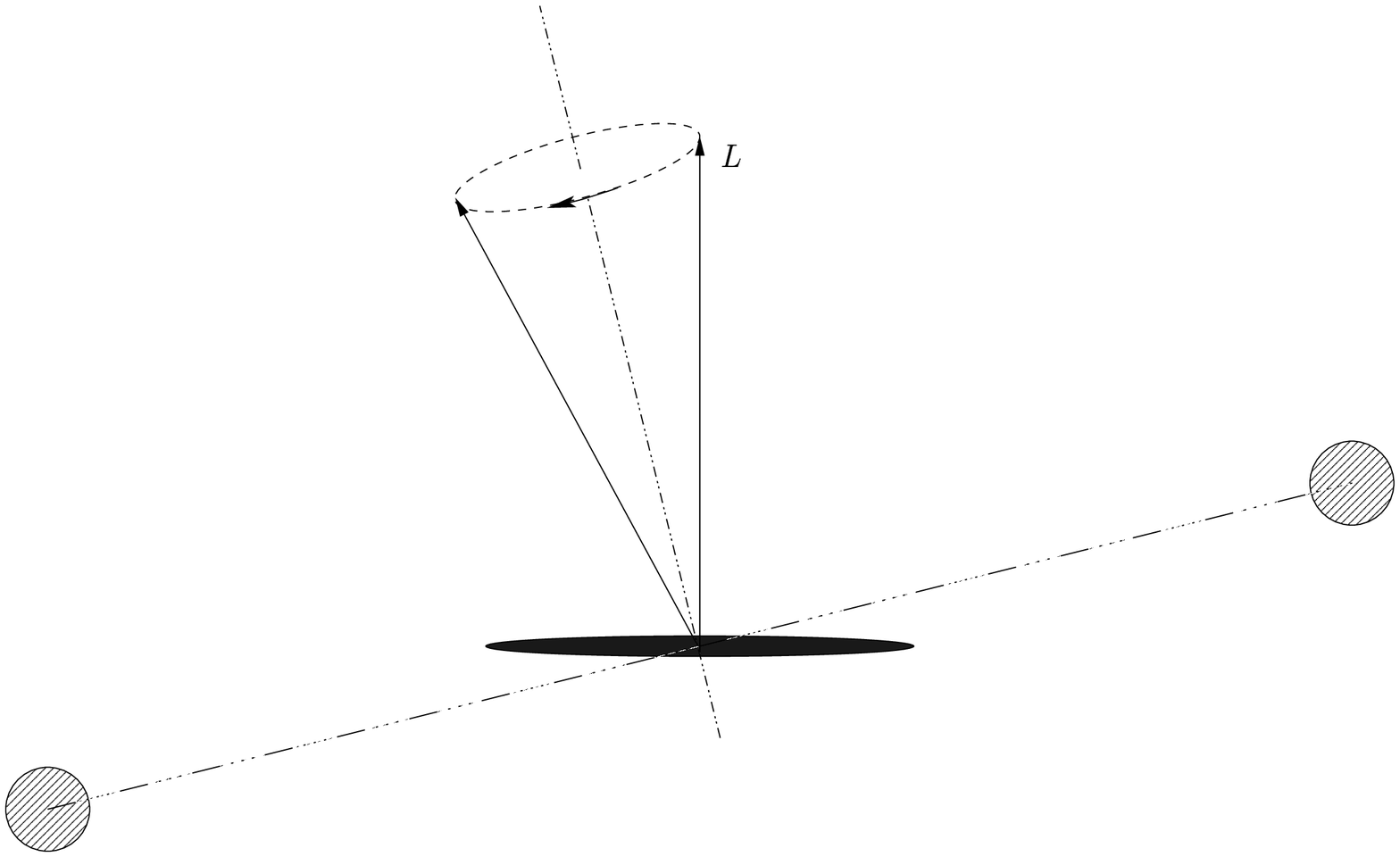}}
\caption[The retrograde precession of the angular momentum of a disc
annulus under the torque of the accreting torus.]
{The retrograde precession of the angular momentum of a disc annulus
under the torque of the accreting torus. The precession is about the
symmetry axis of the torus.}
\label{fig:precess}
\end{figure}

The disc is composed of many concentric rings of massless particles,
which precess at fixed inclination in a retrograde fashion under the
torque of the distant accreting torus.  The precession is illustrated
in Figure~\ref{fig:precess}.

We use the following notation: $\omegap$ is the nodal precession rate
of a narrow annulus of the disc at radius $R$; $\Theta$ is the
angle between the symmetry axis of this annulus and the $z$-axis (\ie,
the normal to the initial unperturbed disc); $\Philon$ is the
azimuthal angle of the symmetry axis of the annulus projected in the
$x$-$y$ plane, which is the plane of the initial unperturbed disc.

Each annulus precesses through an angle
\begin{equation}
\dphi= \int \omegap(t) \; dt,
\label{eqn:dphi}
\end{equation}
where the precession rate due to the quadrupole term only
(Equation~\ref{eqn:omegap_2nd}) is
\begin{equation}
\label{eqn:prec_rate_torus}
\omegapt = \frac{3}{4} \frac{G\Mt R}{\Rt^3 \Vc} \cos i \propto R
\;\;\;{\rm for\;\; } \Vc\sim {\rm const}.
\end{equation}
Clearly $\dphi$ increases with $R$.

From Equations~(\ref{eqn:sinht}) and (\ref{eqn:philon_app}), we find
\begin{equation}
\Theta \approx 2i \; \sin (\dphi/2) \qquad \hbox{for small }
\Theta,\;\; i
\end{equation}
and
\begin{equation}
\Philon \approx - \pi/2 - \dphi /2 \qquad \hbox{ for small }  i  
\label{eqn:philona}
\end{equation}

\begin{figure}
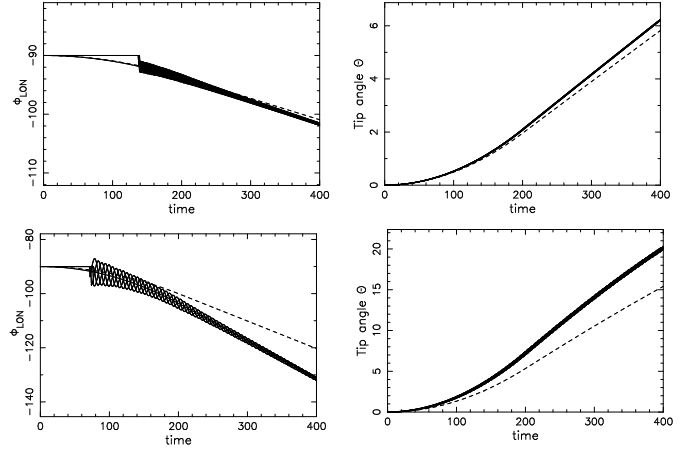
 

\centerline{\includegraphics[angle=-90,
width=.5\hsize]{figs/696_hack4_phi_t_ring_r_3.ps} \hspace{0.02\hsize}
\includegraphics[angle=-90,
width=.5\hsize]{figs/696_hack4_theta_t_ring_r_3.ps}}
\vspace{.02\hsize} \centerline{\includegraphics[angle=-90,
width=.5\hsize]{figs/696_hack4_phi_t_ring_r_7.ps}
\hspace{0.03\hsize} \includegraphics[angle=-90,
width=.5\hsize]{figs/696_hack4_theta_t_ring_r_7.ps}}
\caption[The angles ($\Philon$ and $\Theta$) measured from the model
simulation match the theoretical values very well.]
{The angles $\Philon$ (left) and $\Theta$ (right) measured from the
test particle disc discussed in \S\ref{sec:toymodel}.  Particles at
$r=3.15$ are shown in the top row and at $r=7.05$ in the bottom row.
The light solid lines are the angles measured from 8 particles at each
radius; the heavy solid lines are predictions based on the
fourth-order approximation of $\omegap$
(Equation~\ref{eqn:omegap_total}); the dashed lines are results based
on the second-order approximation (Equation~\ref{eqn:omegap_2nd}).
The fourth-order approximations are almost invisible because they
overlap the curves measured from the 8 particles, indicating excellent
agreement, while the second-order approximations are clearly
inadequate at large radii.  Note that $\Philon$ has been set to
$-90^\circ$ for $\Theta<1^\circ$, to avoid large scatter at early
times.}
\label{fig:anglesmatch}
\end{figure}

For test particles at a given radius, the tip angle $\Theta$ and
$\Philon$ can be measured from the orientation of their angular
momentum vectors.  The angles, $\Philon$ and $\Theta$, measured from
particles at two distinct radii are shown in
Figure~\ref{fig:anglesmatch}; each particle in an annulus has its own
pair of angles at each instant, which vary systematically with phase
giving rise to the spread.  The mean is in qualitative agreement with
these second-order predictions, shown by the dashed curves while the
fourth order (Equation~\ref{eqn:omegap_total}) predictions (solid
curves) agree very well.  Since $\Philon$ can have any value when
$\Theta \simeq 0$, we set $\Philon$ to $-90^\circ$ for
$\Theta<1^\circ$, to suppress noise in these figures.

\begin{figure} 
\centerline{\includegraphics[angle=-90, width=.9\hsize]{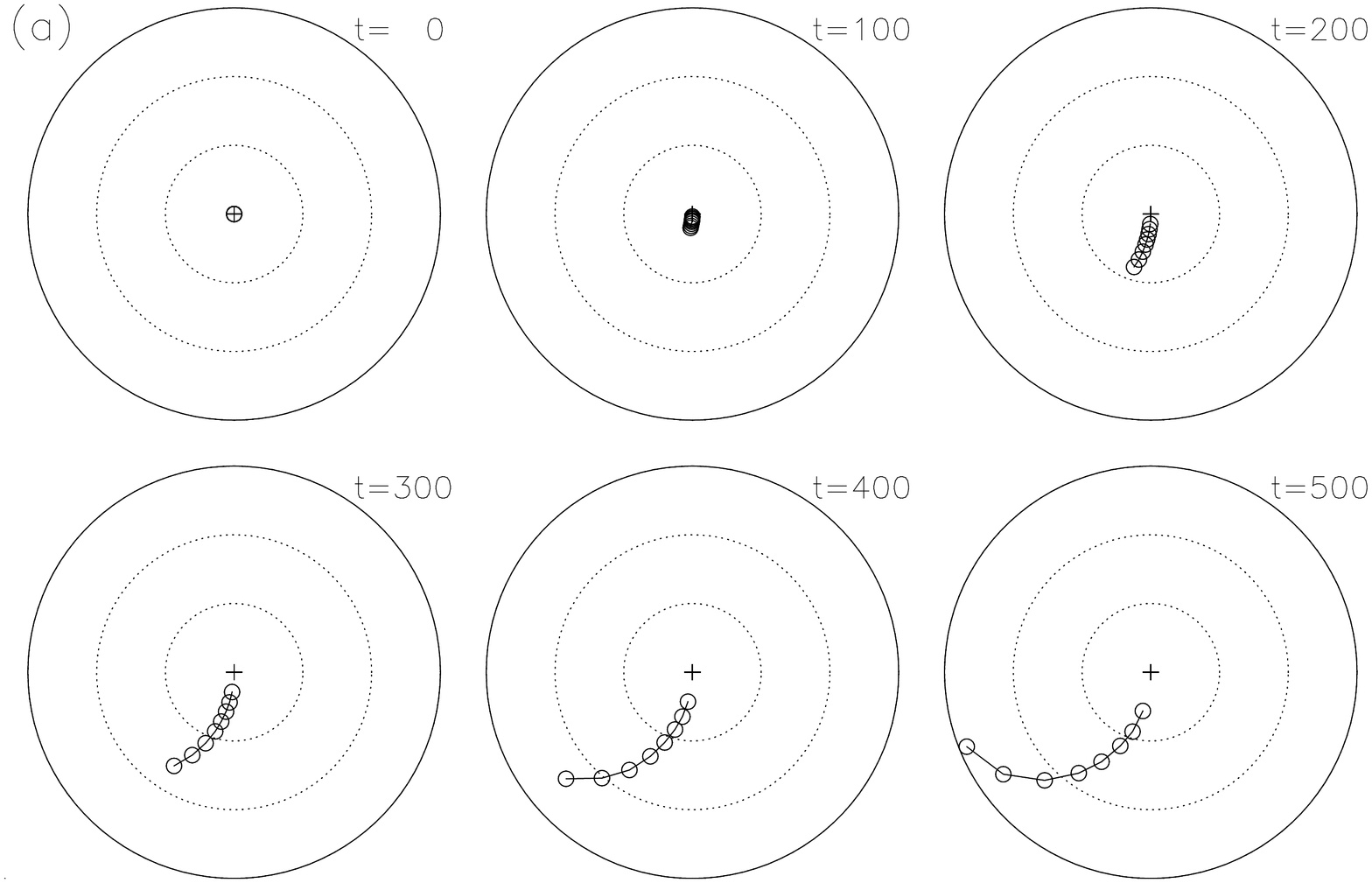}}
\vspace{.05\hsize}
\centerline{\includegraphics[angle=-90, width=.9\hsize]{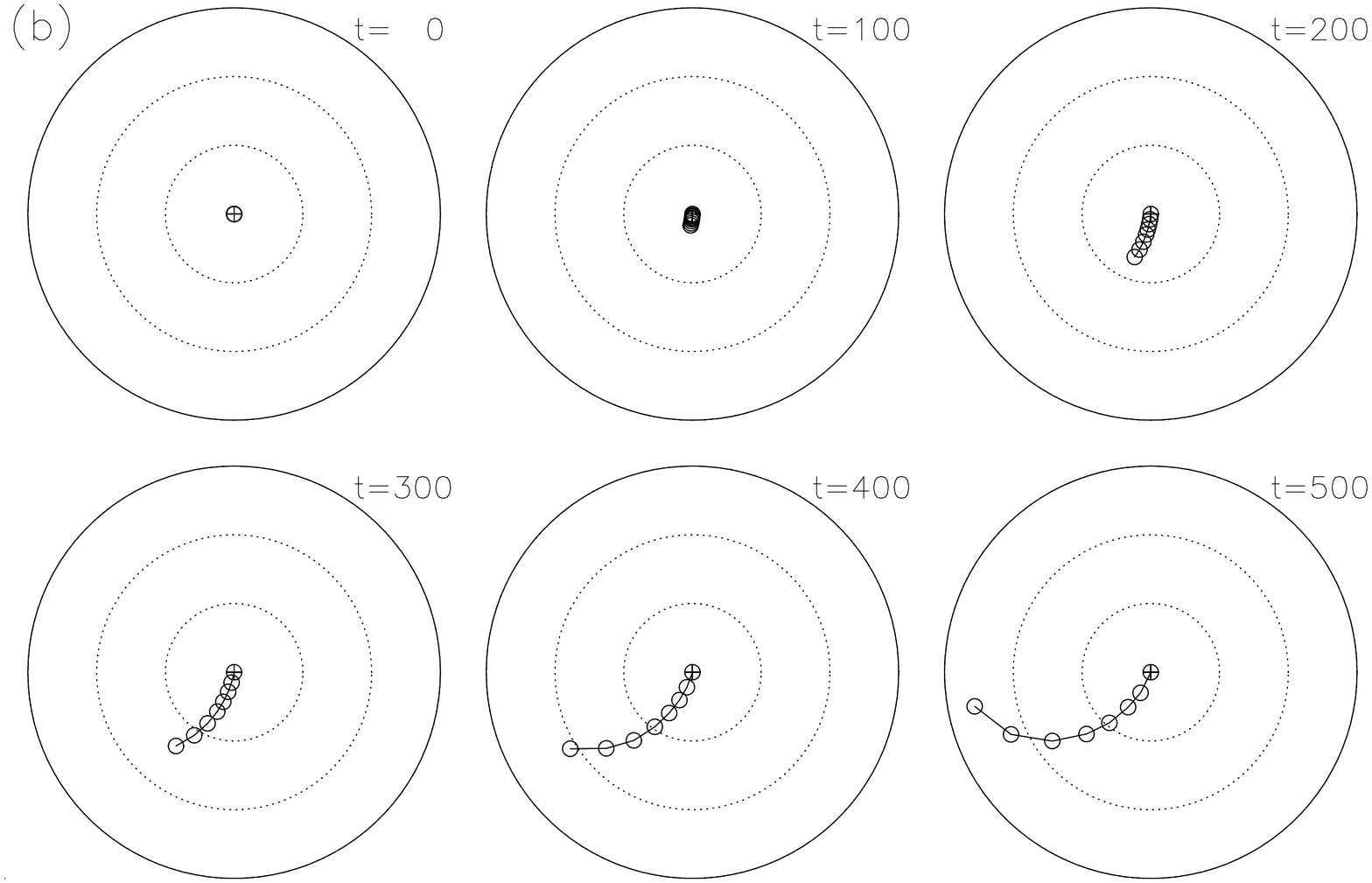}}
\caption[The tip-LON plots of the model in different reference
frames. (a) tip-LON plots with respect to the fixed coordinate system;
(b) tip-LON plots transformed into the inner disc frame of any given
time.]
{Tip-LON plots of the test particle disc model in different reference
frames. (a) The top {\em two} rows show $\Theta$ and $\Philon$ with
respect to the fixed coordinate system of
Figure~\ref{fig:anglesmatch}. (b) The bottom {\em two} rows show the
angles $\theta$ and $\philon$ relative to the inner disc frame at each
time, which is more appropriate for comparisons with observations.
The radial binning scheme is the same as that used in
Figure~\ref{fig:typicallon}, but the concentric dotted circles are
drawn at intervals of $10^\circ$.}
\label{fig:no_disk_tip_lon}
\end{figure}

The increasing rate of retrograde precession with radius causes the
warp angle $\theta$ and $\philon$ to increase outwards also.  We
observe this behaviour in our test-particle simulation as shown in
Figure~\ref{fig:no_disk_tip_lon}.  In this simplified model, the warp
appears as the result of the differential precession and the LON forms
a trailing spiral.

This simulation also serves as a nice check of parts of our numerical
code: forces from the torus particles and the time integration of the
test particle disc are calculated accurately.

\subsection{Including disc self-gravity}
\label{sec:varymass}
We separate a self-gravitating disc into an inner disc and an outer
disc.  The dense inner disc, which contains most of the disc mass,
tilts rigidly as a whole; it is strongly coupled together due both to
self-gravity and to radial velocity spread of stars, which communicate
stress across the disc via epicyclic excursions
\citep{deb_sel_99}. The gradual locking of the disc due to self-gravity
was also found in
\citet{lovela_98}.  The low-density outer disc, on the other hand, is
much less cohesive because self-gravity is weak and epicycles are
small; the outer disc behaves more as a collection of test particles
therefore.

As the torus mass rises, the whole disc starts to precess as described
in \S\ref{sec:toymodel}.  The torque from the torus causes the inner
disc, which is strongly cohesive, to precess slowly as a whole in a
retrograde manner about the symmetry axis of the torus, while the
outer disc precesses more rapidly.  The developing misalignment between
the inner and outer disc causes the particles in the outer disc to feel an
additional torque from the massive inner disc.  The rate of precession
of the outer disc due to the torque from the inner disc follows a
different rule from that due to the torus.

The disc potential at larger radii can be approximated as the sum of a
monopole and a quadrupole term \citep[Equation~{6-84}]{bin_tre_87},
\begin{equation}
\Phi_{\rm d} \approx -\frac{G\Md}{r}+\frac{G}{2r^3} \Ptwo \cdot 
\int_0^\infty 2 \pi {R^{\prime}}^3 \Sigma(R^{\prime}) dR^{\prime}.
\end{equation}
Using equation~(\ref{eqn:expdisk}) for an exponential disc, we have
\begin{equation}
\Phi_{\rm d} \approx -\frac{G\Md}{r}+\frac{G\Md}{r} \frac{3\Rd^2}{r^2} \Ptwo.
\end{equation}
In the same manner as for the torus \citep[Appendix~\ref{app:pre}, see
also][]{kah_wol_59, gunn_79}, we find the precession rate of
test rings due to the disc is
\begin{equation}
\omegapd \sim \left(\frac{3}{2} \cdot \frac{3\Rd^2}{r^2} \cdot
\frac{G\Md}{r}\right)(r \Vc)^{-1} \propto r^{-4} \quad{\rm for }\quad
\Vc\sim {\rm const}.
\label{eqn:prec_rate_disk}
\end{equation}

\begin{figure} 
\centerline{\includegraphics[angle=-90, width=.9\hsize]{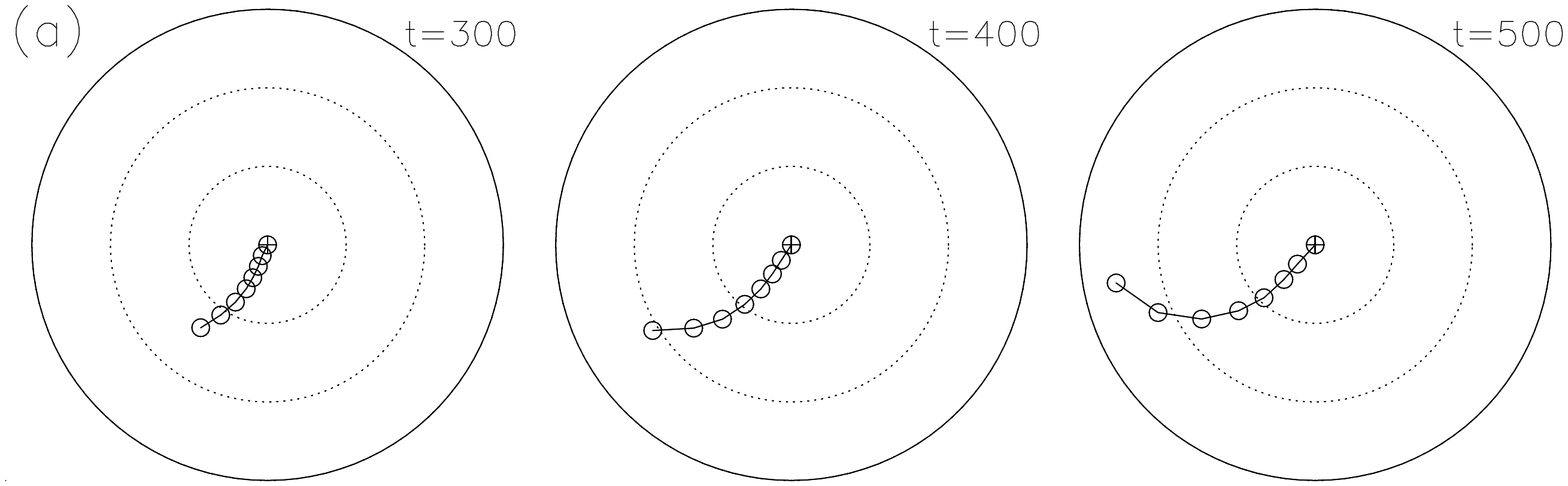}}
\vspace{.05\hsize}
\centerline{\includegraphics[angle=-90, width=.9\hsize]{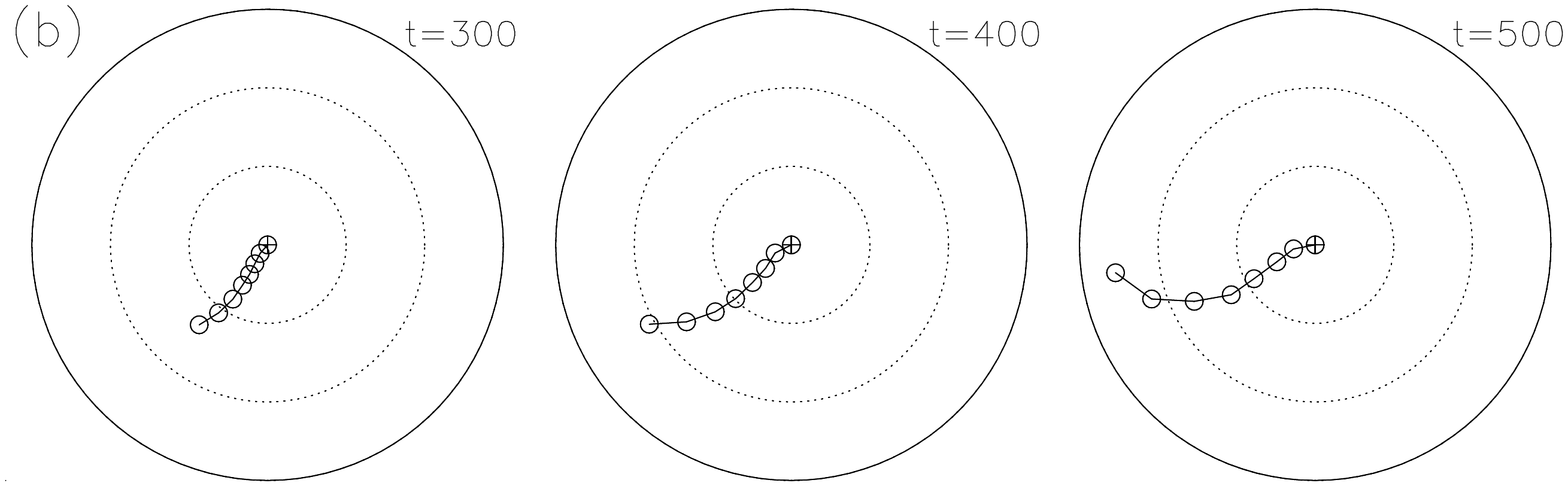}}
\vspace{.05\hsize}
\centerline{\includegraphics[angle=-90, width=.9\hsize]{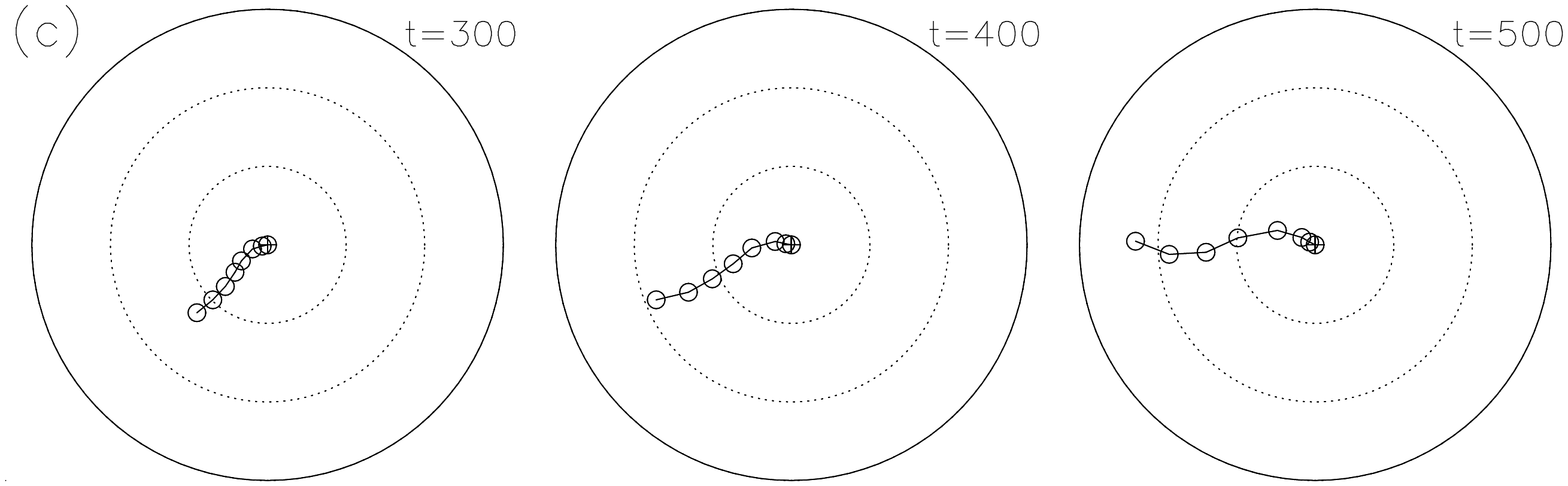}}
\vspace{.05\hsize}
\centerline{\includegraphics[angle=-90, width=.9\hsize]{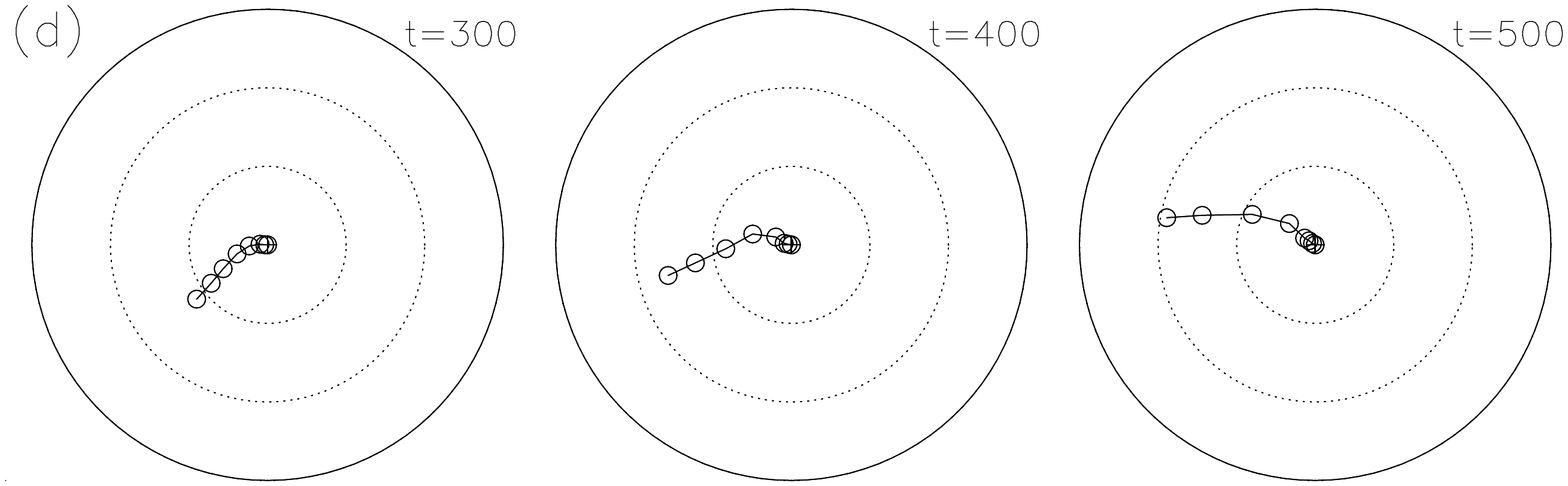}}
\caption[Tip-LON plots as we gradually increase the strength of the
disc from $\Md=0.1$ to $\Md=2.0$.]
{Each row shows part of the warp evolution for different disc masses:
(a) $\Md=0.1$, (b) $\Md=0.3$, (c) $\Md=1.0$, and (d) $\Md=2.0$.  Note
the $\Md=0$ case is shown in Figure~\ref{fig:no_disk_tip_lon}(b).  The
spiral formed by the inner LON clearly changes from trailing to
leading as $\Md$ is increased.  Dotted circles mark $10^\circ$
intervals.}
\label{fig:tip_lon_add_disk}
\end{figure}

While still producing retrograde precession, the rate from this
additional forcing term varies with radius in the opposite sense of
that from the torus.  In the case of forcing by the inner disc, the
precession rate Equation~(\ref{eqn:prec_rate_disk}) decreases with
increasing radius, which in isolation would tend to make the LON
develop a leading spiral.

The outer disc is now subject to two torques, from the torus and from
the inner disc, both of which cause retrograde precession, but at rates
that vary radially in opposite senses.  Whether the resulting spiral
in the warp LON should lead or trail over the range of interest
depends on relative magnitudes of the torques, and is best determined
from simulations.

We have therefore performed a series of simulations with increasingly
massive discs to show that the torque from the inner disc dominates in
our canonical simulation, which develops a leading spiral in the LON
(see Figure~\ref{fig:typicallon}). Figure~\ref{fig:tip_lon_add_disk}
presents the tip-LON diagrams from four experiments with increasingly
larger disc masses (from $\Md=0.1$ to $2.0$) to demonstrate the effect
of the disc mass.  These experiments continue to employ the same
spherical rigid halo and torus parameters as above.  The general trend
confirms that the leading spiral gradually dominates the inner LON as
$\Md$ is increased; the LON has become completely leading in
Figure~\ref{fig:tip_lon_add_disk}(d) where $\Md=2.0$.  (Note that the
result for $\Md=1$ in a rigid halo differs from that in
Figure~\ref{fig:typicallon} where the halo is responsive.)

\subsection{A test particle disc in a live halo} 
\label{sec:experiment2}
The rigid spherical halo in the above experiments, cannot respond to
the precessing disc, and therefore does not affect how the LON curves.
A halo composed of live particles should acquire a distorted shape in
response to the fields of the torus and the disc.  These effects are
both included in a fully self-consistent simulation. Here we study
these effects separately and demonstrate that the live halo does not
alter the conclusions we reached in \S\ref{sec:toymodel} and
\ref{sec:varymass}.

\begin{figure} 
\centerline{\includegraphics[angle=-90, width=.9\hsize]{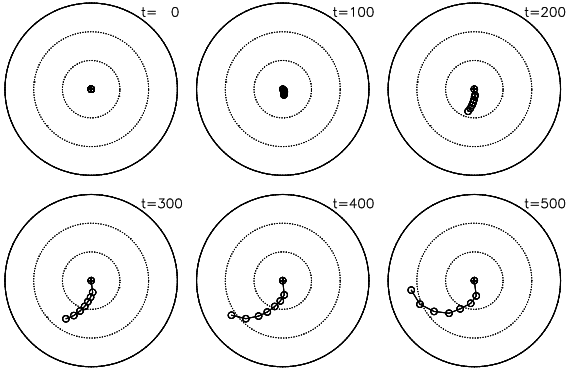}}
\caption[The evolution of the tip-LON plots for a test particle disc in a
live halo.]
{The evolution of the warp for a test particle disc in a live
halo. The only difference from Figure~\ref{fig:no_disk_tip_lon} is
that the live halo becomes slightly oblate under the influence of the
distant torus. }
\label{fig:tip_lon_749}
\end{figure}

In a further experiment, we replace the rigid halo in the model of
\S\ref{sec:toymodel} with a live one with nearly the same
potential.  Figure~\ref{fig:tip_lon_749} shows that in this experiment
the LON becomes trailing, as we found in \S\ref{sec:toymodel} where
the halo is spherically rigid.  This again supports our argument that
the massive disc, instead of other factors like a live halo, is the
primary reason why the LON forms a leading spiral for warps induced by
cosmic infall. The small differences in the angle of precession
between Figures~\ref{fig:tip_lon_749} and \ref{fig:no_disk_tip_lon}(b)
indicate that the oblateness of the halo (induced by the torus)
contributes only slightly to the precession of the disc.

We discuss the response of a live halo with a massive disc in
\S\ref{sec:haloresponse}.

\begin{figure} 
\centerline{
\includegraphics[angle=0, width=.45\hsize]{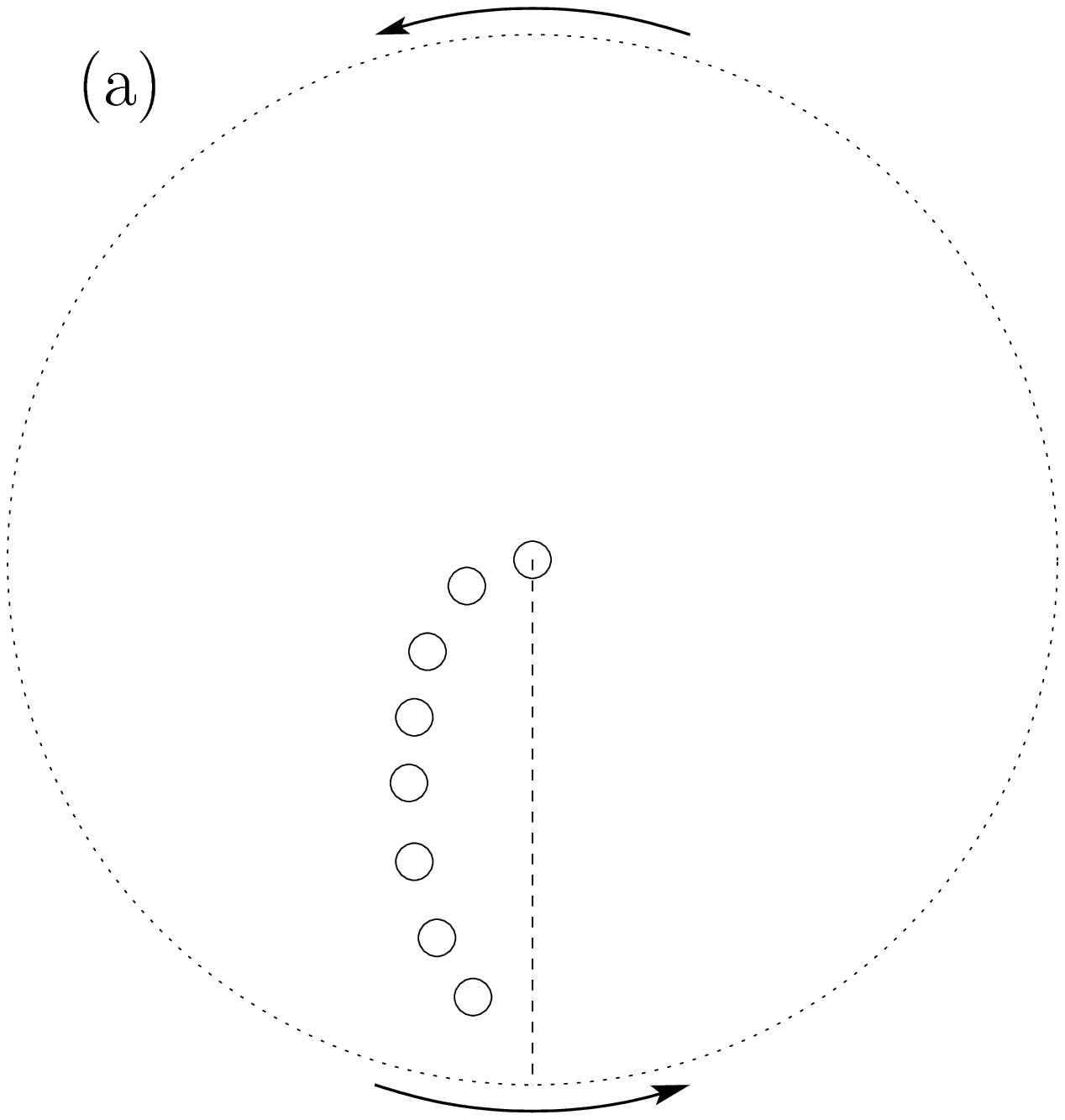}
\includegraphics[angle=0, width=.45\hsize]{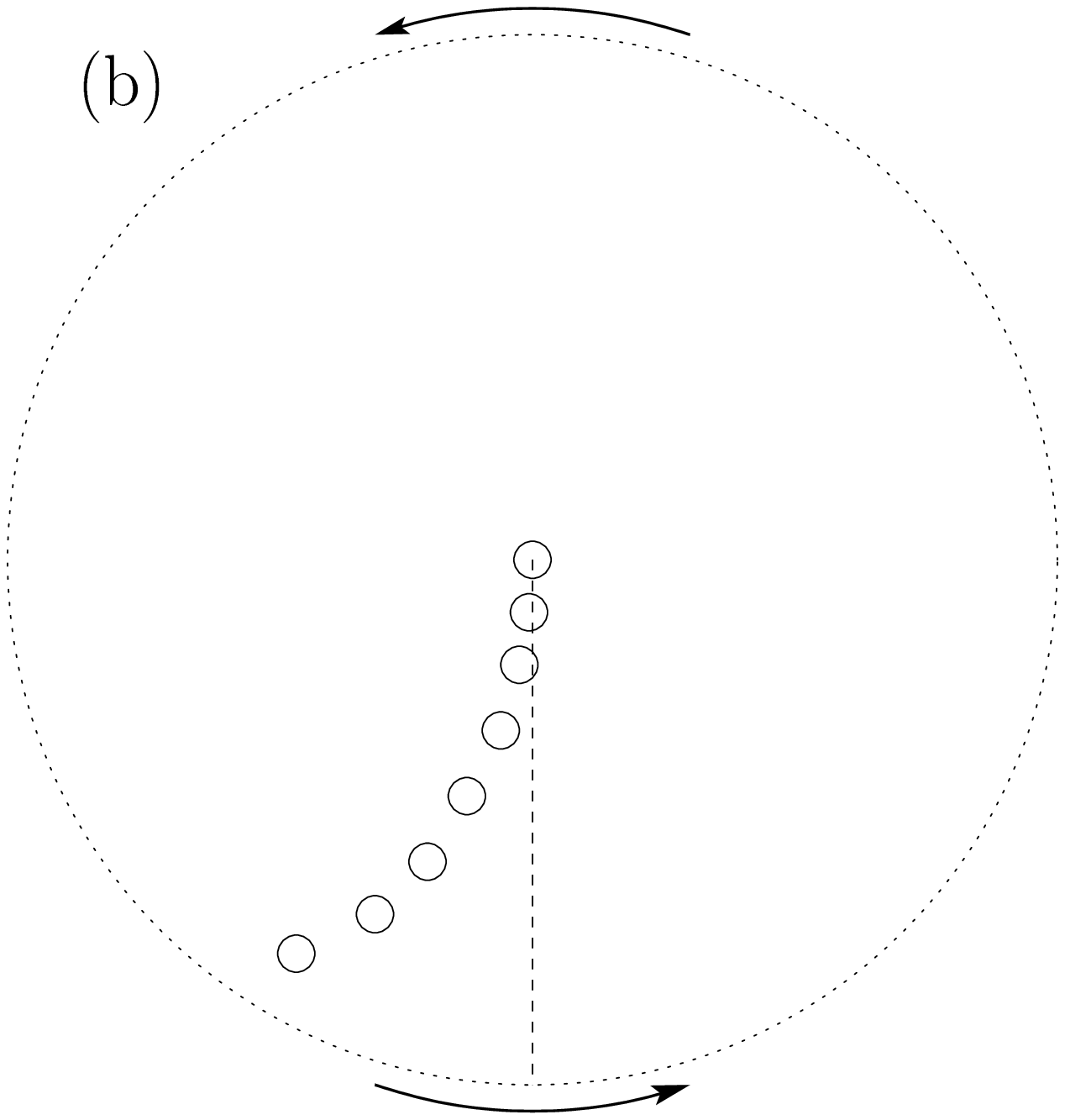}
}
\caption[Schematic illustration on the spirality of the LON of
warps.]
{Schematic illustration showing two possible spiralities of the warp
LON.  (a) A leading spiral caused by a precession rate $\omegap$ that
is a monotonically decreasing function of $R$ such as $\omegap
\sim \omegapd \propto R^{-4}$; (b) A trailing spiral resulting from a
precession rate $\omegap$ that is a monotonically increasing function
of $R$ such as $\omegap \sim \omegapt \propto R$. The arrows outside
the plots indicate the direction of galaxy rotation.}
\label{fig:lon_illus}
\end{figure}

\begin{figure} 
\centerline{\includegraphics[angle=-90, width=0.9\hsize]{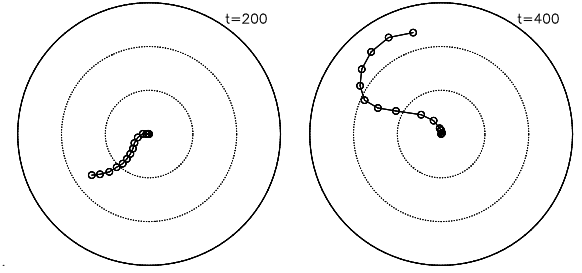}}
\caption[The spirality of LON twists from leading to trailing at very
large radius in the canonical simulation.]
{The spirality of LON twists from leading to trailing at very large
radius (the transition occurs near the eighth point from the centre,
\ie, the annulus centered on $R=7.55\Rd$).  These tip-LON plots are
almost same as those at $t=200$ and $400$ in
Figure~\ref{fig:typicallon}, except that 14 radial bins are used to
cover all particles, including the additional massless ones; the
binning scheme is similar to that described in
\S\ref{sec:leadinglon}.  Note the disc rotates in a counter-clockwise
sense.}
\label{fig:lontwist}
\end{figure}

\subsection{Spirality twist of LON at large radius}
\label{sec:twist}
In Figure~\ref{fig:tip_lon_add_disk}(c), we can clearly see the
spirality of LON twists from leading to trailing in the tip-LON plot
at $t=500$.  Here we show that such spirality twist is a generic
feature for warps caused by an external quadrupole field giving
outward-increasing rate of retrograde precession.

As shown above, the sense of spirality of the LON is determined by the
radial variation of the precession rate $\omegap$, and is measured by
$\Philon$ in the tip-LON plots.  From Equation~\ref{eqn:dphi} and
\ref{eqn:philona}, we have
\begin{equation}
\Philon \approx {\rm const} - \frac{1}{2} \int \omegap(R, \;t) \; dt
\end{equation}
the negative sign indicates that the precession is retrograde.
Whether the spiral leads or trails depends on whether $\omegap$ falls
or rises as $R$ increases, as illustrated in
Figure~\ref{fig:lon_illus}.

We have argued that the torque due to the massive inner disc, which
causes $\omegap \sim \omegapd \propto R^{-4}$ to dominate, causing the
leading spiral found in our simulations.  Beyond some critical radius,
$\Rtw$, precession is dictated by the accreting torus, $\omegap \sim
\omegapt \propto R$, and a trailing LON should be expected.  The
critical radius $\Rtw$ can be estimated as follows: The magnitude
ratio of the torques exerted on a test ring at radius $R$, due to the
torus (the first term of Equation~\ref{eqn:torque_2nd} as the
second-order approximation) and to the massive inner disc
(Equation~\ref{eqn:prec_rate_disk}), is approximately
\begin{equation}
\frac{\tau_{\rm torus}}{\tau_{\rm disc}} \sim 
\left(\frac{3}{4} \frac{G\Mt}{\Rt^3} R^2\right)
\left(\frac{9}{2} \frac{G\Md}{R}\frac{\Rd^2}{R^2} \right)^{-1}
= \frac{1}{6} \frac{\Mt}{\Md} \frac{R^5}{\Rd^2 \Rt^3}.
\end{equation}
The radius of the spirality twist is approximately where
\begin{equation}
\left[ \frac{\tau_{\rm torus}}{\tau_{\rm disc}}\right ] _{R=\Rtw}  \sim 1.
\end{equation}
So we have
\begin{equation}
\Rtw \sim \left[ 6 \left( \frac{\Md}{\Mt} \right) \right] ^{1/5} \cdot
\left[\Rd^2 \Rt^3 \right]^{1/5}.
\end{equation}
For our torus parameters ($\Mt = 2.5$, $\Rt = 15$), we find $\Rtw \sim
6.1\Rd$.  In Figure~\ref{fig:tip_lon_add_disk}(c) where $\Md=1.0$, the
spirality twist occurs near the sixth point from the centre (\ie, near
the annulus centered on $R=6.15\Rd$), so it agrees well with the
expected $\Rtw$.

$\Rtw$ can also be understood in the context of a Laplacian surface,
where the net torque vanishes \citep[chap.~6.6]{bin_tre_87}.  It is
easy to show that $\Rtw$ is also the transition radius of the
Laplacian surface; for $R \ll \Rtw$, the Laplacian surface almost
coincides with the inner disc, which dominates the precession for
small radii, while for $R \gg \Rtw$, the Laplacian surface follows the
equatorial plane of the torus, which dominates the precession at large
radii.

The inner part of a responsive halo becomes oblate due to the massive
inner disc.  The co-aligned oblate inner halo adds significantly to
the torque from the inner disc, causing $\Rtw$ to be larger than
estimated above, to perhaps $7\Rd \la \Rtw \la 8\Rd$.  So the LON
spirality twist could occur at radii $\ga 2\RHo$.

Thus the spirality twist can be seen only if there is a tracer at very
large radii.  We do not see the spirality twist in
Figure~\ref{fig:typicallon} because very few particles lie beyond
$\sim 8\Rd$.  To reveal the expected spirality twist, we add into our
canonical simulation 20,000 massless test particles in the radial
range of $3\Rd < R < 12\Rd$, to mimic the outer gas layer.  We used 14
radial bins to cover all particles; the binning scheme is similar to
that described in \S\ref{sec:leadinglon}.  The spirality twist shows
up in Figure~\ref{fig:lontwist} near the seventh from outermost point
(\ie, near the annulus centered on $R=7.55\Rd$).  As expected, the
spirality twist grows more pronounced with time.

Note that a twist can occur only if the forcing causes an
outward-increasing rate of retrograde precession, as for our
simplified torus.  More realistic perturbations may not have this
effect.

\section{The halo response}
\label{sec:haloresponse}
In this section, we return to our canonical simulation
(\S\ref{sec:results}) with the live halo and the massive disc
($\Md=1.0$).  

\subsection{Disc--inner halo alignment}
\begin{figure} 
\centerline{\includegraphics[angle=-90, width=.9\hsize]{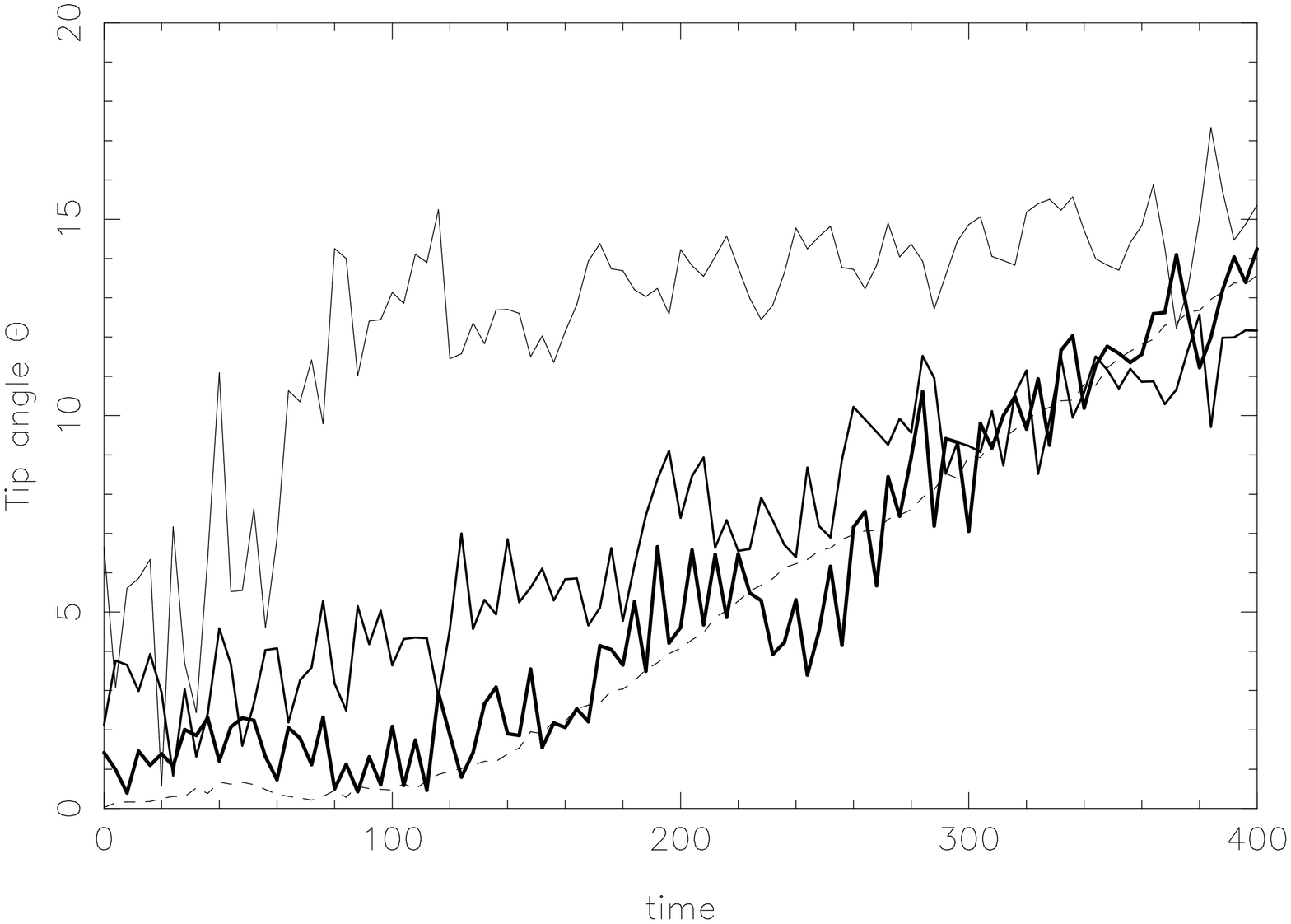}}
\vspace{.04 \hsize}
\centerline{\includegraphics[angle=-90, width=.9\hsize]{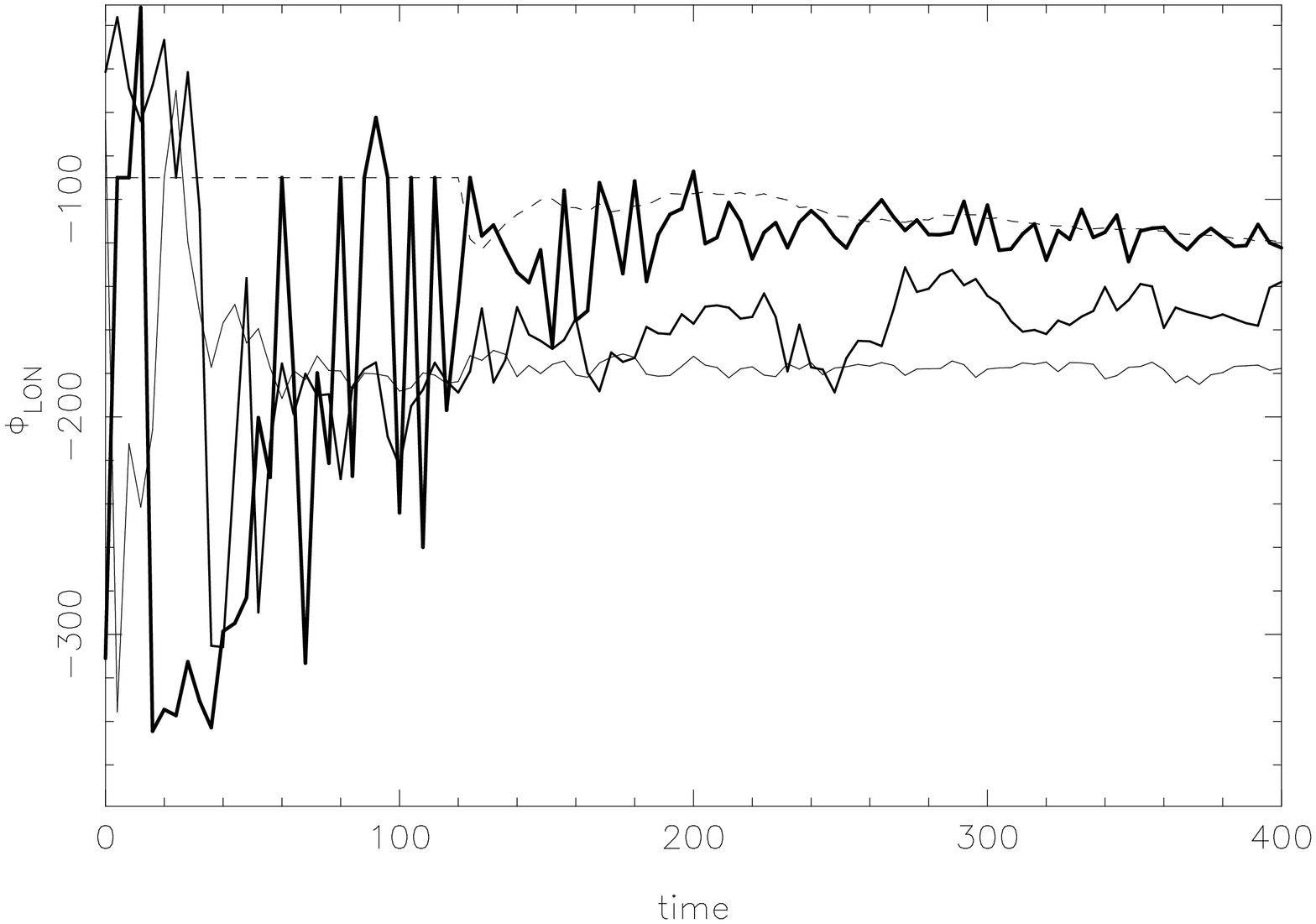}}
\caption[The response of the halo in the process of warp
formation. (a) The time evolution of the tip angle $\Theta$; (b) The
time evolution of $\Philon$.]
{The time evolution of the halo axes in the canonical run, (a) the tip
angle $\Theta$ and (b) $\Philon$ both measured in the fixed initial
frame.  The solid curves, with decreasing thickness, represent the
orientations of the inner halo ($0 < r < 4\Rd$), the central halo
($4\Rd < r < 8\Rd$) and the outer halo ($8\Rd < r < 12\Rd$)
respectively, while the dashed curves show these angles for the inner
disc.}
\label{fig:halores}
\end{figure}

Figure~\ref{fig:halores} shows the orientation of the
halo in three spherical bins ($0 < r< 4\Rd$, $4\Rd < r < 8\Rd$, and
$8\Rd < r < 12\Rd$) measured from the canonical simulation.  The inner
halo (heavy line) follows the inner disc (dashed) closely, while the
outer halo (light line) quickly aligns with the outer torus, which has
a tip angle $\Theta \sim 15^\circ$ and azimuth $\Philon \sim
180^\circ$.  The halo orientation in the middle bin is intermediate.
It is worth stressing that the halo mass interior to $5\Rd$ is almost
five times that of the disc, yet its orientation does not follow that
of the outer halo and massive torus, but instead follows that of the
lighter disc, which precesses around the axis of the torus because of
its angular momentum.  Because the massive inner halo is oblate and
aligned with the inner disc, it adds significantly to the precession
and curving of the LON in the outer disc. \citet{bin_etal_98} also
found that the inner halo quickly realigns with the disc, even when a
large initial misalignment is imposed.

\begin{figure}
\centerline{\includegraphics[angle=-90,
width=.9\hsize]{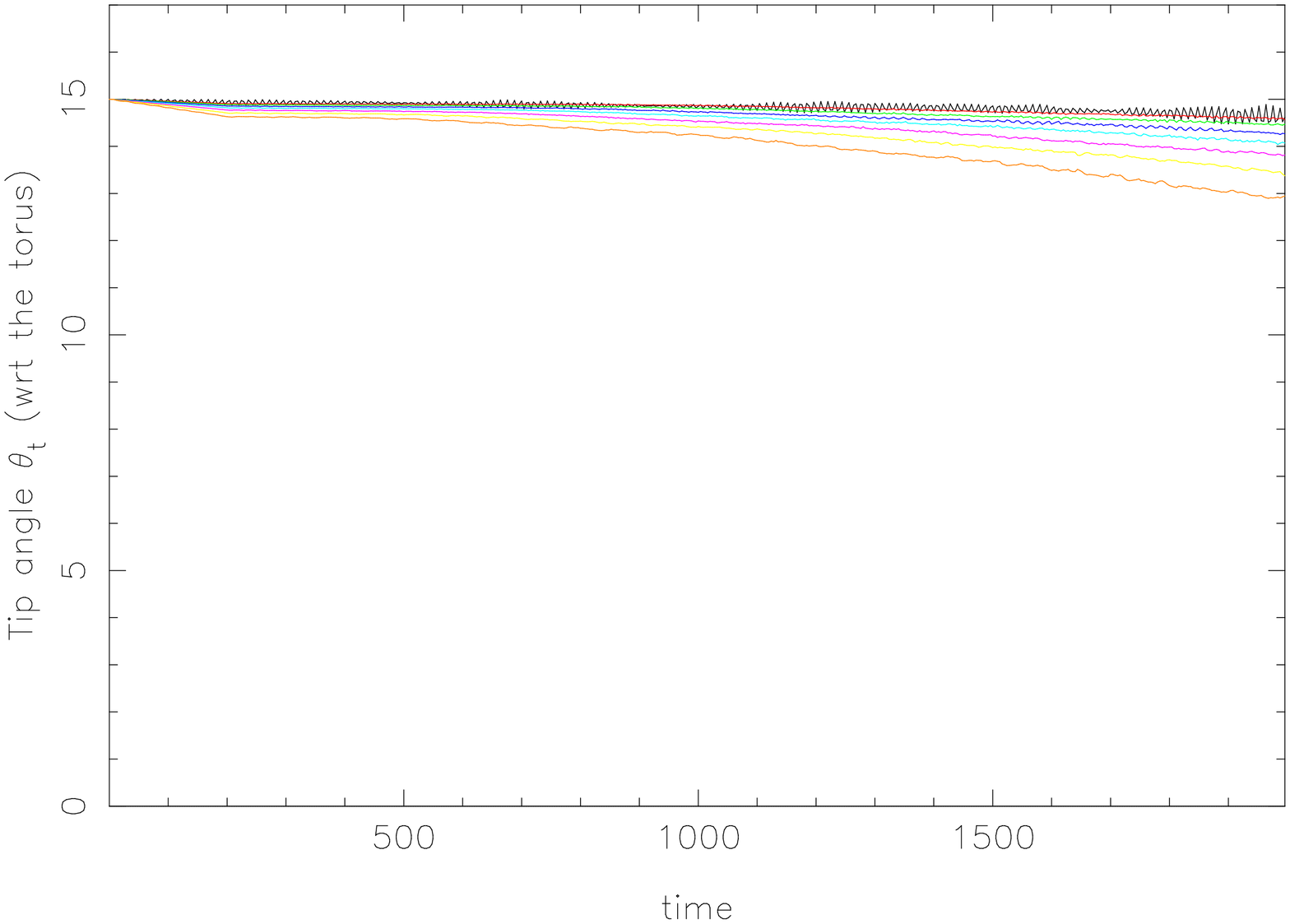}}
\vspace{.04 \hsize}
\centerline{\includegraphics[angle=-90,
width=.9\hsize]{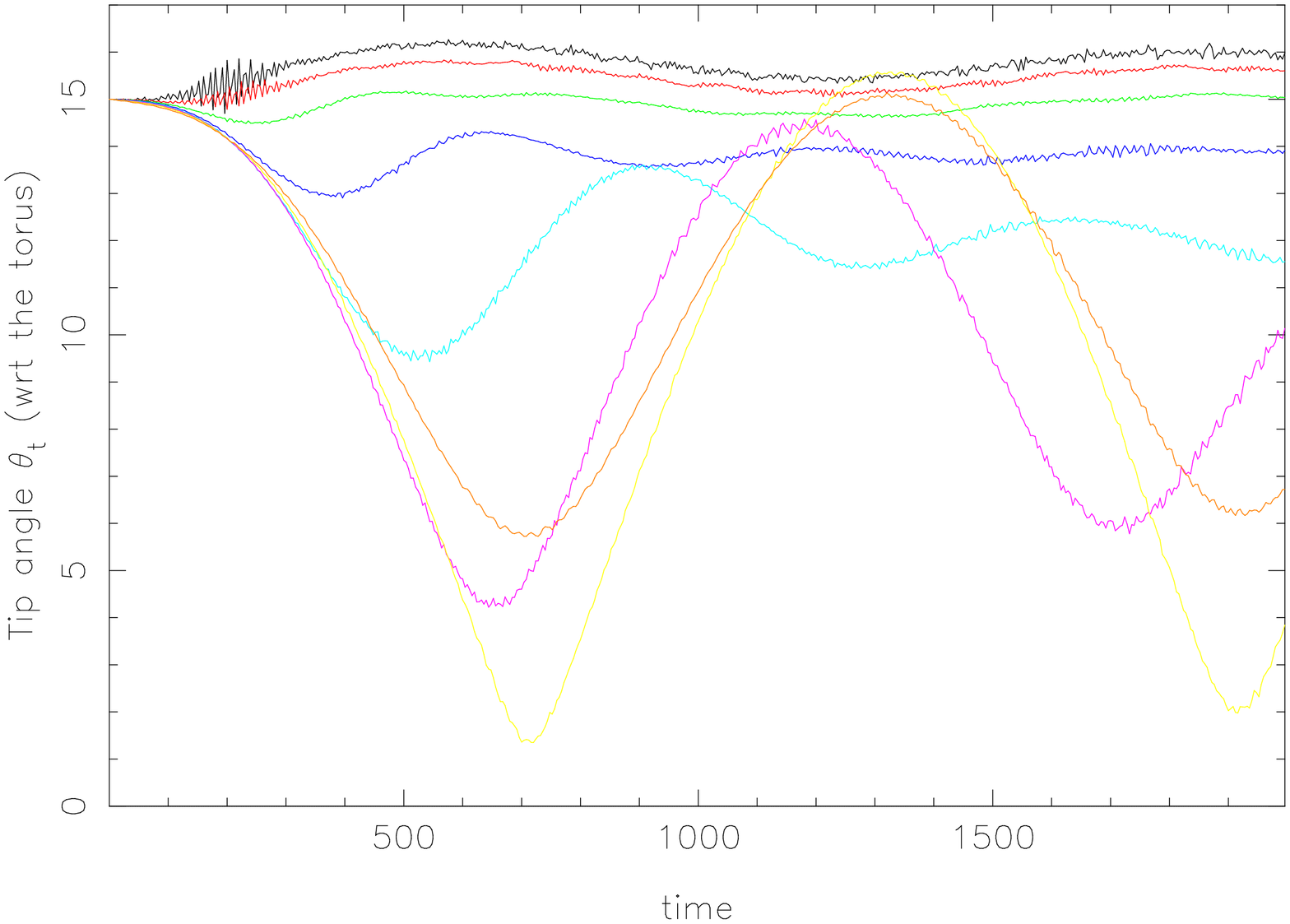}}
\vspace{.04 \hsize}
\centerline{\includegraphics[angle=-90,
width=.9\hsize]{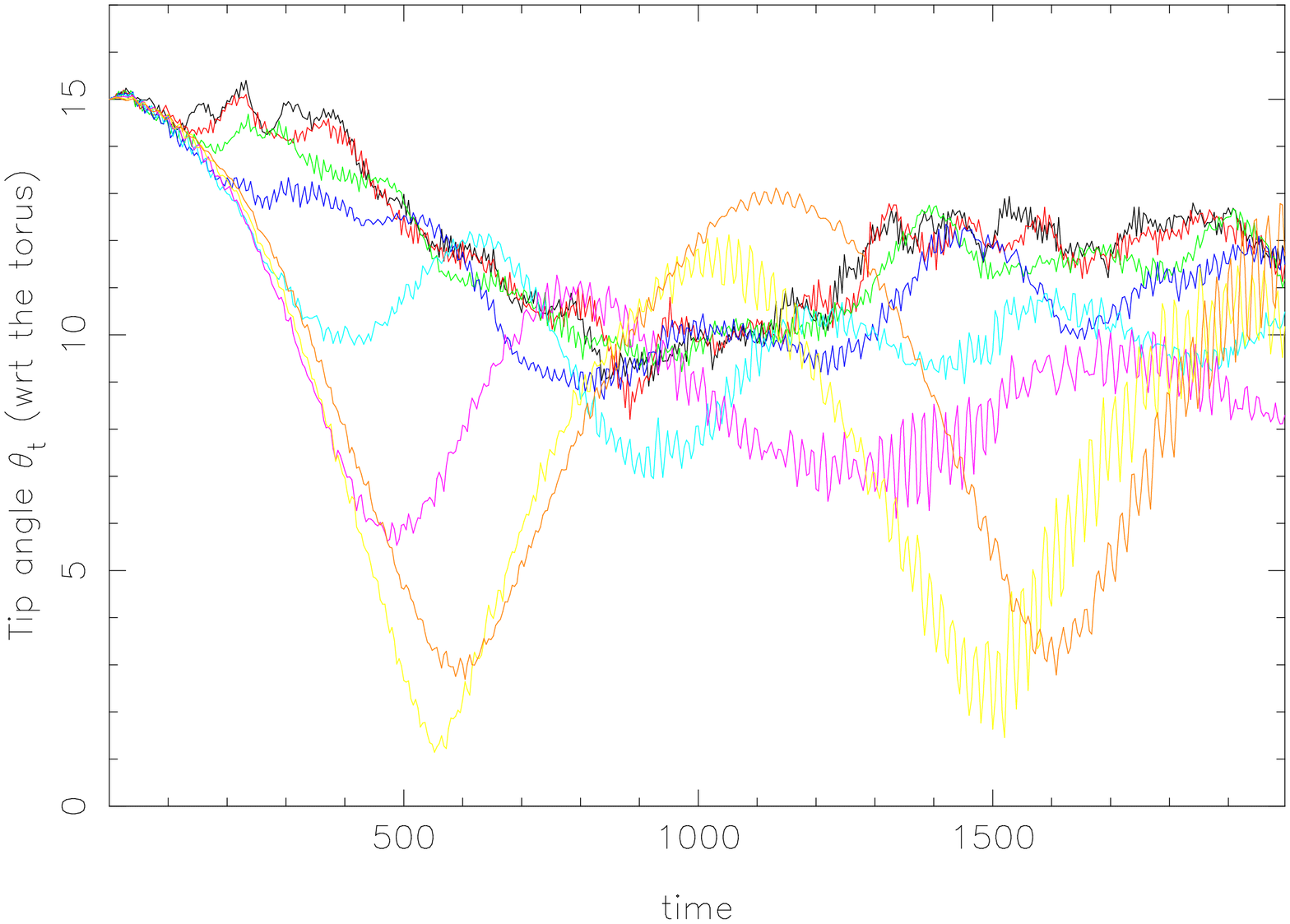}}
\caption[(a) The time evolution of $\theta_t$, the tip angle with
respect to the torus plane; (b) As for (a), but for a massive,
responsive disc in a rigid halo (Run CR); (c) As for (a), but for a
massive disc in a responsive halo (the canonical Run CL).]
{(a) The time evolution of $\theta_t$, the tip angle with respect to
the torus plane, of the inner disc ($0 < R < 3\Rd$, the black line)
and 7 equal-width annuli spanning the radial range $3\Rd < R < 7.9\Rd$
(represented by the red, green, blue, cyan, magenta, yellow, and
orange lines with increasing radius, respectively) for a rigid halo +
test particle disc run.  (b) As for (a), but for a massive, responsive
disc in a rigid halo (Run CR).  (c) As for (a), but for a massive disc
in a responsive halo (the canonical Run CL).  The slight deviations
from $15^\circ$ at later times in (a) are caused merely by averaging
over a finite radial range.}
\label{fig:halodamp}
\end{figure}

\subsection{Halo damping}
Figure \ref{fig:halodamp} shows that halo damping of the warp by the
live halo is weak. Now for the convenience of description, we name the
canonical run as ``Run CL'' (\ie\ canonical run {\it live} halo). We
introduce ``Run CR'' (\ie\ canonical run {\it rigid} halo) with the
almost same halo potential\footnote{Note that the two potentials
cannot be exactly the same, because the live halo is slightly oblate
due to the gravity of the disc initially, and its oblateness evolves
with time.} as in the canonical run except that the halo is rigid and
stays exactly spherical.  Panel (a) shows that nested rings of test
particles on circular orbits in a rigid halo precess at constant
inclination, $\theta_t$, to the plane of the torus.  Since rings
precess at different rates, the apparent deviations from $15^\circ$ at
later times are caused by averaging rings with the same inclinations,
yet different azimuths, over a finite radial range. (We have verified
that the inclinations of individual particles with respect to the
torus plane indeed remain close to $15^\circ$, as expected.)  Panel
(b) shows that self-gravity in a massive disc causes differential
precession about the axis of the inner disc (Run CR). The
quasi-periodic modulation of the inclinations of the outer rings in
(b) is a manifestation of the extra torque from the inner disc. The
initial decrease of the inclinations of most annuli to the torus plane
(between $t=0$ and 600) has nothing to do with dynamical friction or
halo damping since the halo is rigid in this case.  The final panel,
(c), shows the effect of using a live halo (the canonical Run CL).
The precession due to the disc is augmented by the aligned flattening
of the inner halo, which is partly responsible for the differences
from panel (b), but a full explanation of all the details in this
Figure is complicated; \eg, we would need to take into account the
inclination of the halo at intermediate radii.  The important result
is that even after 2000 dynamical times, the mean inclination of the
inner disc to the torus has not decreased by more than $3^\circ$; \ie,
damping from the live halo is very weak.

Our result differs from the conclusions of \citet{dub_kui_95} and
\citet{nel_tre_95} because the scenarios studied are different.  Their
models have a very large initial misalignment between the halo and the
inner disc and a significant precession rate calculated according to
the ``modified tilt mode'' in \citet{spa_cas_88}. Another difference
is that the halos in their models were given no time to adjust to the
strong and rapidly-precessing warp initially inserted in.

In our simulations, the precession rate of the inner disc is very low,
and the inner halo follows its motion closely, as shown in
Figure~\ref{fig:halores}, which is why dynamical friction between the
two components is negligible.  Only the outer halo is misaligned and
damping comes mainly from coupling between the inner and outer halo.
Such coupling is weak for the isotropic halo models we used
\citep{nel_tre_95}. This is also consistent with the recent
cosmological hydrodynamical simulations by \citet{bai_etal_05}, which
found that the relative orientations of inner ($r < 0.1r_{\rm vir}$)
and outer ($r > 0.1r_{\rm vir}$) halo are uncorrelated.

\section{Discussion}
\label{sec:discussion}

\subsection{The source of the external torque}
\label{sec:discuss_external_torque}

The torque in our model arises from a uniform, massive torus centered
on the galaxy and inclined to the disc, as originally adopted by
\citet{jia_bin_99}.  The warp is driven by the quadrupole field
of the torus, which varies as $\Mt/\Rt^3$, and therefore a less
massive torus at a smaller radius has a similar effect, as we have
verified in other experiments.  However, a weaker perturbation would
give rise to a milder warp that would take longer to form.

A natural direct interpretation of the torus is an idealized
representation of the stream of stars and dark matter from a disrupted
orbiting satellite, if the mass is sufficiently large.  There are
numerous examples of accreted satellites, or companion galaxies
orbiting in planes that are misaligned with that of the disc of the
host galaxy.  The Sgr Dwarf and the Magellanic Clouds are clear
examples in the Milky Way, but their masses are generally on the low
side unless their halos are substantial.

On the other hand, the torque from the torus is similar to that of a
flattened outer halo, which may also have misaligned angular momentum
\citep{qui_bin_92}.  Conceptually, one could separate the halo into
two mass components: an outer halo flattened in a plane misaligned
with the disc and having a uniform, quasi-spherical core, and the
remainder -- an inner halo with any arbitrary radial mass profile,
that may be strongly peaked to the centre of the galaxy.  Regardless
of how flattened the inner halo may be, it exerts no torque on the
inner disc, since it is generally aligned with the disc plane (see
\S\ref{sec:haloresponse}).  In this picture, the gravitational
potential of the outer oblate halo alone might approximate the form:
$\Phi_{\rm OH}(R, z) = v_0^2 / 2 \cdot \ln \left(R_c^2 + R^2 +
z^2/q_\Phi^2 \right)$, which arises from a flattened isothermal
envelope with a quasi-spherical, uniform core of radius $\sim R_c$,
with a density scale set by $v_0$.  The ellipticity of the
equipotential surfaces $\epsilon_\Phi = 1 - q_\Phi$ is approximately
$\epsilon_\rho/3$, where $\epsilon_\rho$ is the ellipticity of the
halo density.  It is easy to show the torque due to this outer halo
exerted on a ring in the disc at radius $r \!<\! R_c$ is
$\langle\tau(r)\rangle \sim {r^2 v_0^2} (q_\Phi^{-2}-1)/(2R_c^2)$,
which has the same radial dependence as the torque of a uniform
torus. The effects discussed in \S\ref{sec:twist} and
\S\ref{sec:warppersist} are dependent on the outward-increasing rate
of retrograde precession of our model. The torque from our adopted
torus is of the same order of magnitude as such an outer halo with a
typical flattening in the range $0.6 < q_\rho < 0.9$ depending on
$R_c$.

Cosmological simulations that include baryonic infall have not yet
settled on a set of robust predictions for the shape and alignment of
the outer halo.  A misaligned potential of the form just described is
highly idealized, but has served as a useful theoretical exercise to
understand warps.  It has yielded some insight that may capture
features of the torques that drive real warps.

\subsection{Persistence of warps}
\label{sec:warppersist}
The windup rate of a warp depends on the radial variation of the net
precession rate. (The amount of the windup can be measured by $\cot i
= \Delta t R \cdot d\omegap/dR$, see \citealt{bin_tre_87}.)  Forcing
by our adopted torus causes differential precession of the opposite
sign from that of the disc, making the total retrograde precession
less differential, especially in the warp region ($4\Rd<R<10\Rd$).
Thus the stiff inner disc and the two precessions work together to
slow the windup of the warp.

However, it is more interesting to study the evolution of a pre-warped
massive disc without any external forcing.  We therefore conducted two
further experiments in which we removed the forcing torus after some
evolution.  We ran cases with both live (CL--torus) and rigid
(CR--torus) halos, derived from the canonical runs CL (live halo) and
CR (rigid halo).  In both cases, we gradually removed all the torus
particles in a linear fashion between $t=400$ and $600$ after which,
no torus particles remain.  (Note that in all runs the torus mass was
grown linearly to its maximum between $t=0$ and 200.)

In these two runs, a warp was already well developed by $t=400$.
Ramping down the torque from the torus allows us to study the unforced
behaviour of the warped disc.  In run CR--torus, the disc is affected by
its self-gravity only (because the halo is rigidly spherical), but in
run CL--torus, the effect of self-gravity of the live halo is also
included.

\begin{figure}
\centerline{
\includegraphics[angle=-90,width=.45\hsize]{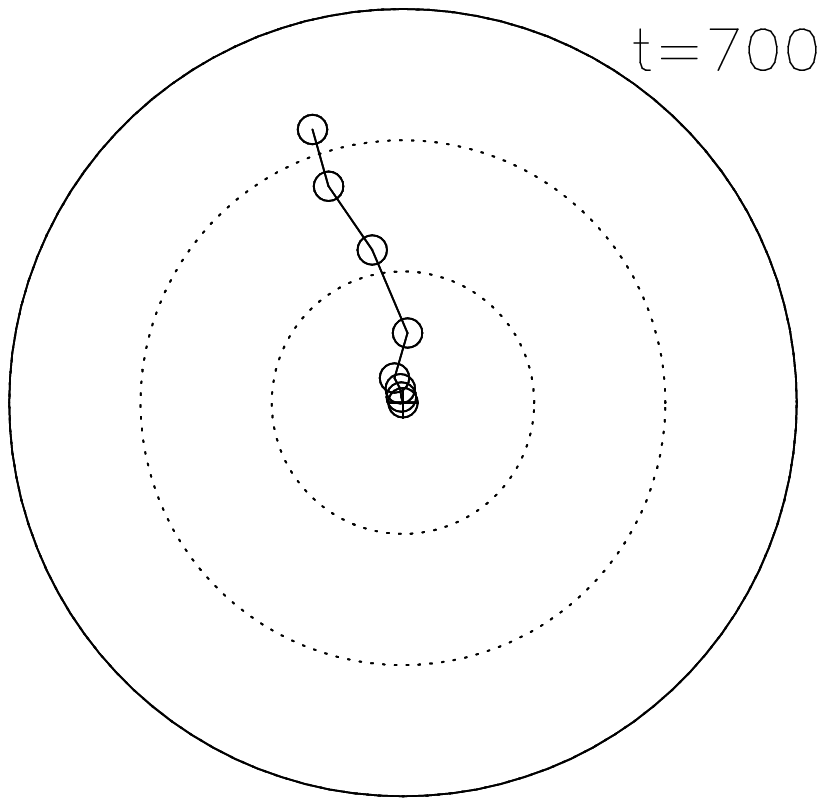}
\includegraphics[angle=-90,width=.45\hsize]{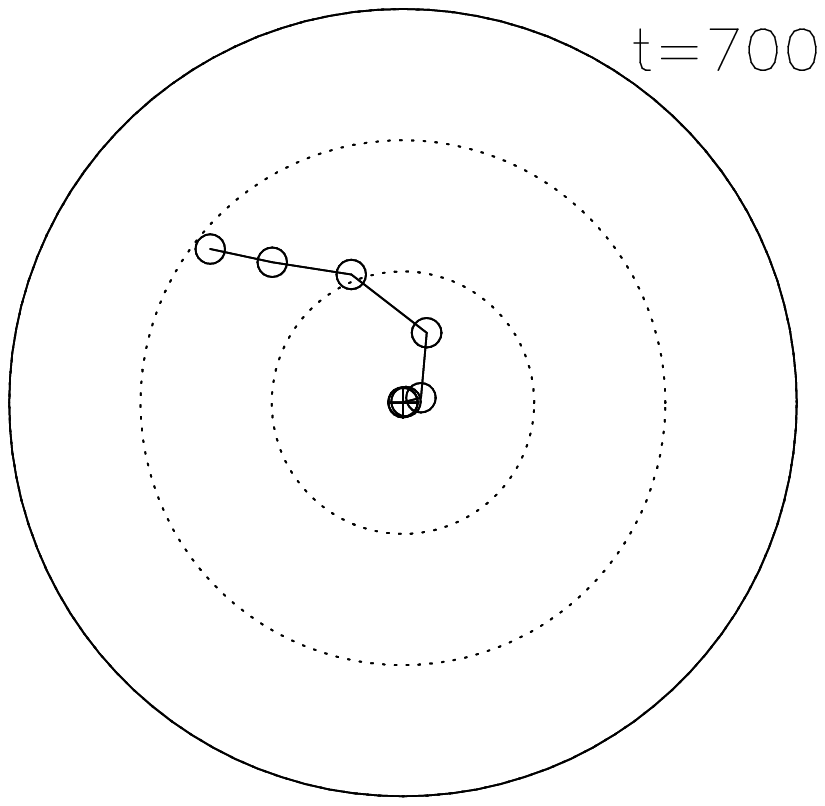}}
\caption[The torque from the torus helps to alleviate the windup of
the warp --- the case of a rigid halo. (a) The LON of Run CR at
$t=700$; (b) The LON of Run CR--torus at $t=700$] {Removing the
external forcing by the torus has little effect on the amplitude of
the warp, but does affect the windup rate.  The LON of Run CR (left) at
$t=700$ is slightly leading, but almost straight while that for Run
CR--torus (right) at the same time is more wound up.  Note that
these runs, with a rigid halo, had the exactly same LON at $t=400$.}
\label{fig:windup_rigid}
\end{figure}

\begin{figure}
\centerline{
\includegraphics[angle=-90,width=.45\hsize]{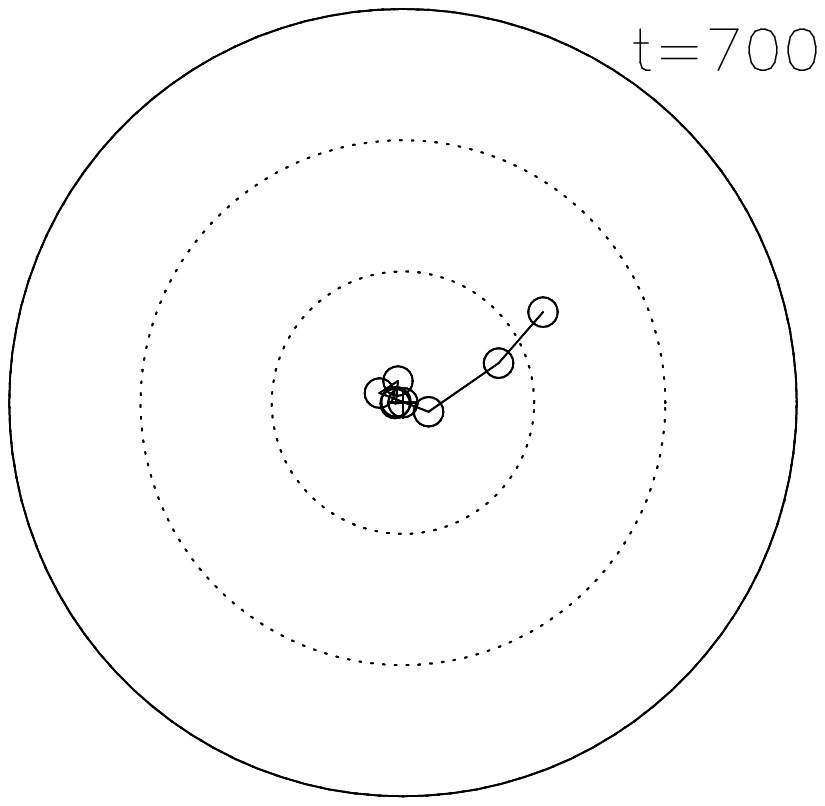}
\includegraphics[angle=-90,width=.45\hsize]{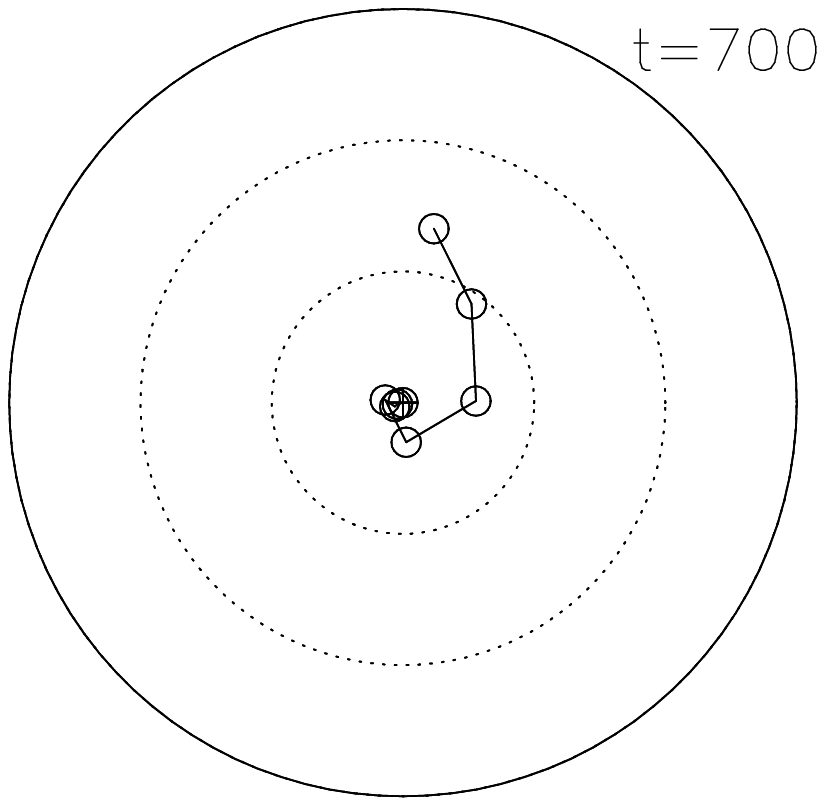}}
\caption[The torque from the torus helps to alleviate the windup of
the warp --- the case of a live halo. (a) The LON of Run CL (the
canonical run) at $t=700$; (b) The LON of Run CL--torus at $t=700$]
{As Figure~\ref{fig:windup_rigid} but for a live halo.  The tip-LON
plot from of Run CL (the canonical run) at $t=700$, shown on the left,
again has similar amplitude to that for Run CL--torus on the
right but is less wound.}
\label{fig:windup_live}
\end{figure}

In Figure~\ref{fig:windup_rigid} and Figure~\ref{fig:windup_live} we
compare the LON between runs in which the torus was left in place and
the cases in which it was removed.  Figure~\ref{fig:windup_rigid}
shows the two cases at $t=700$ for the rigid halo, and
Figure~\ref{fig:windup_live} shows the corresponding two cases with
the live halo.  Note that both experiments in each pair are identical
at $t=400$.  Two results are evident from this comparison, regardless
of whether the halo is rigid or live: first, the amplitude of the warp
is little changed by the removal of the forcing torus confirming that
damping from the halo is weak; second, the LON in runs with the torus
removed is more wound than in runs where the torus remains.  This
second point confirms that the torque from the torus is important to
balance out the differential precession due to the stiff inner disc,
and reduces the windup rate of the warp, but the extra winding is mild
even $\sim3$ Gyr after the torus starts to be taken away.  Thus the
warp can persist for a few Gyrs after the imposed quadrupolar
perturbation is removed.

\subsection{Observations of the LON at large radius}

Our analysis has shown that the LON of warps in the inner region forms
a leading spiral due to the potential of the disc and inner halo.  The
fact that the LONs of most warps form leading spirals over an extended
radial range may also imply that the disc mass is significant in the
central region of these galaxies (e.g. Figure~\ref{fig:tip_lon_add_disk}).

The spirality of the LON at larger radii, on the other hand, contains
information on the mass distribution in the outer halo that drives the
warp.  For example, if warps are formed by external perturbations that
exert torques of the form of our adopted torus, the LON should twist
from leading to trailing at some large radius.  The absence of a twist
would imply a perturbation of a different form.

We have looked for signs of a twist in the largest sample of galaxies
with tip-LON data to date \citep{briggs_90}.  The transition radius
$\Rtw \sim 8\Rd$ for the torus we study (\S\ref{sec:twist}), a value
that has a mild dependence on the torus properties ($\Rtw \sim
[\Mt/\Rt^3]^{-0.2}$).  If most of the large, late-type spirals in
Briggs's sample have Holmberg radii in the range $\RHo=R_{\rm 26.5}
\sim 4.4 \Rd$, we expect $\Rtw \sim 2\RHo$ and we therefore examine
those having LON data beyond $2\RHo$.

Six galaxies (NGC 628, NGC 5055, NGC 2841, NGC 1058, NGC 3344 and M83)
in this sample have LON data outside $2\RHo$.  We find that NGC 628
and NGC 5055 show signs of the LON spirality twist near $2\RHo$; NGC
2841 and NGC 1058 show weak signs of the spirality twist; the LON of
M83 is quite straight near $2\RHo$, and might have spirality twist at
an even larger radius; the LON of NGC 3344 is very noisy, but the
spiral of the overall LON trend does not seem to twist.  Thus perhaps
four out of six galaxies reveal hints of spirality twists in their
LONs.

Thus currently available data do not provide convincing evidence for
the spirality twist, mainly because the LON data beyond $2\RHo$
\citep{briggs_90} are generally sparse and noisy.  The outer
morphology of the LON is worth more detailed investigation in deeper
HI 21 cm observations of gas-rich galaxies, since it may contain
valuable information on the shape and profile of the mass distribution
in the outer halo.

\begin{figure}
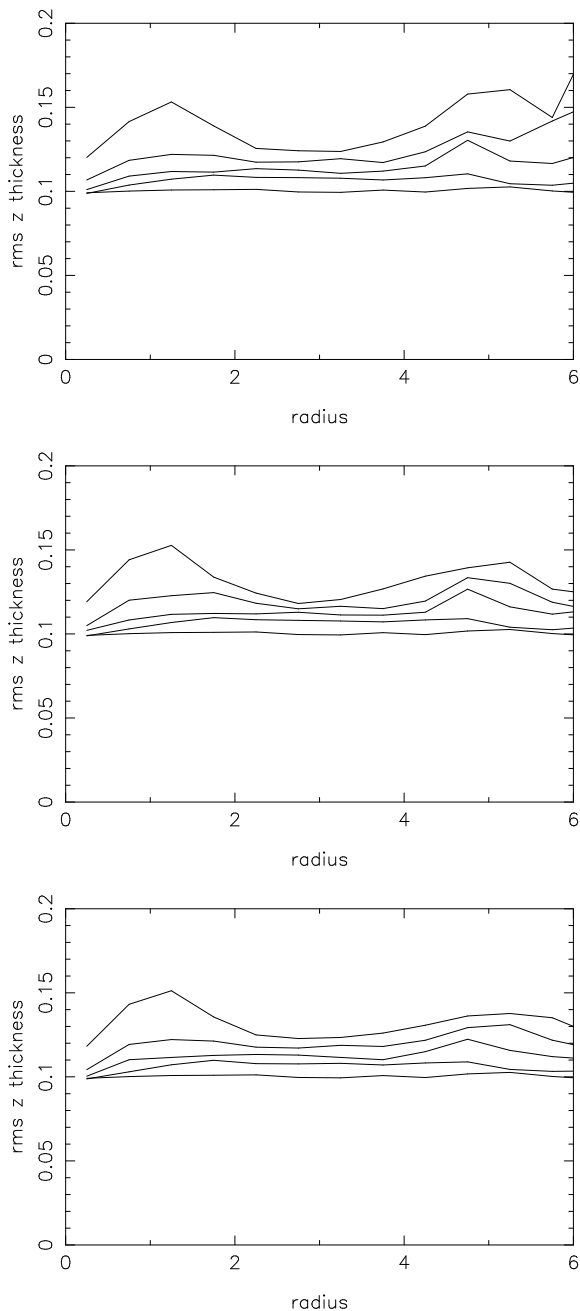

\centerline{\includegraphics[angle=-90,
width=.9\hsize]{figs/zthick_732.ps}}
\vspace{.04 \hsize}
\centerline{\includegraphics[angle=-90,
width=.9\hsize]{figs/zthick_754_torus_0_tilt.ps}}
\vspace{.04 \hsize}
\centerline{\includegraphics[angle=-90,
width=.9\hsize]{figs/zthick_761_no_torus.ps}}
\caption[The evolution of the disc thickness (rms z dispersion)
profiles in various runs.]
{Radial variation of disc thickness (rms z dispersion) relative to the
mean inclination, at different times.  The curves from bottom to top
show values at $t=0$, 100, 200, 300, and 400, respectively.  The
canonical simulation is shown in the top panel.  The middle panel
shows a comparison run with the same torus parameters as the canonical
simulation except that the torus is in the disc plane, and the bottom
panel shows another comparison run without a torus.}
\label{fig:intr_thick}
\end{figure}

\begin{figure}
\centerline{\includegraphics[angle=0,width=\hsize]{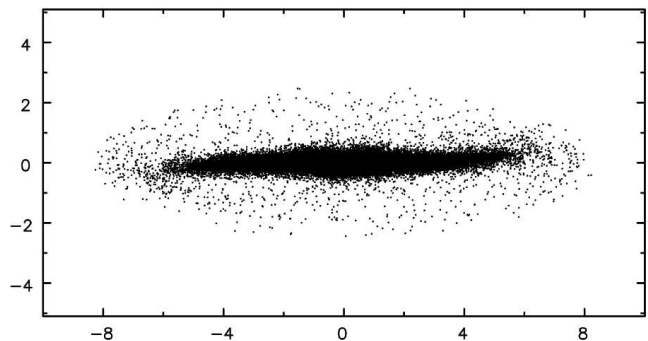}}
\caption[A distant warp may be mistaken as a ``thick disc'' due to the
projection effect.]
{The projected view of the warp in Figure~\ref{fig:morph}a when the
viewpoint is rotated by $90^\circ$ from that in Figure~\ref{fig:morph}a.}
\label{fig:thickdisk}
\end{figure}

\subsection{The apparent thick disc due to warping}
The intrinsic thickness of the disc about its mean inclination is not
greatly increased by warping.  Comparison of the three panels in
Figure~\ref{fig:intr_thick} reveals that the evolution of the disc
thickness is little affected by an aligned torus, and a tilted torus
causes only a slight increase for $R \ga 4\Rd$.

But a warped, luminous disc will contribute some light outside the
thin, inner disc which, from some viewing directions (see
Figures~\ref{fig:thickdisk}), may resemble a low-surface brightness
thick disc.  Suppose an idealized warp that starts from $R_{25}$ at an
angle $\alpha$ to the inner disc and has a straight LON.  (We define
$\alpha$ to be the kink angle between the flat inner disc and the
steepest edge of the warp when seen in projection along the LON, \ie\
it is not the warp angle $\theta$ defined above.)  The projected
surface brightness profile along the minor axis of the edge-on disc
will clearly be exponential with vertical scale-height
$h_z=\Rd\sin\alpha$, and its brightness is roughly $I(R_{\rm
warp}\approx R_{25})/\sin\alpha \sim (25 + \log_{10}\sin\alpha) \muB$.
The angle $\alpha \sim 45^\circ$ for the warp in
Figure~\ref{fig:morph}a and Figures~\ref{fig:thickdisk}.  Thus, if the
warp starts in the luminous disc, it could possibly affect conclusions
about thick discs from photometry of edge-on galaxies.

\section{Conclusions}
\label{sec:conclusions}
Now that the role of a live halo has been properly taken into account,
the idea that galaxy warps are manifestations of long-lived warp modes
appears to have reached a dead end.  But warps as responses to
external forcing remain viable, provided that suitable perturbations
are frequent enough.  Hierarchical galaxy formation scenarios
guarantee late infall within gravitationally bound subgroups, both
diffuse and in lumps, and the infalling matter is unlikely to share
the angular momentum vector of the early disc and halo
\citep{ost_bin_89,qui_bin_92}.

The simulations presented here are motivated by this idea, but are
limited to an in-depth study of one highly idealized case.  We examine
how an initially flat particle-disc with random motion, embedded in a
live halo, gradually acquires a warp as a result of a steady, applied
external torque.  Rather than striving for realism, we have
endeavoured to understand how the warp develops and why the LON
generally forms a loosely-wound, leading spiral.

Despite having focused mainly on one case, we find a long-lived,
large-amplitude warp that resembles those observed in many respects.
The disc is flat in the inner part, and starts to warp at $\sim 5\Rd$
into an open leading spiral, which is consistent with the warping
rules found by \citet{briggs_90}.

The external torque causes the disc to precess in a retrograde sense.
The warp develops because the precession rate of the inner disc, which
is strongly cohesive due to its self-gravity and random motion, is
slower than that of the outer disc.  The growing misalignment between
the inner and outer disc gives rise to a new torque acting on the
outer disc from the massive inner disc, and the inner halo that is
coupled to it.  The torque from the interior mass is responsible for
the leading spiral of the line of nodes -- our adopted external field
alone would produce a trailing spiral.  The fact that the LON of most
warps forms a leading spiral over an extended radial range seems to
imply a massive disc (e.g. Figure~\ref{fig:tip_lon_add_disk}).

On the other hand, the leading spiral tells us little about the cause
of the warp, but such information might be revealed by the behaviour
of the warp at large radii.  We show that the LON twists to trailing
at very large radii from our experiments with a perturbation that
alone would cause the warp to form a trailing spiral.  Thus better
observational data on the shape of the LON at very large radii may
reveal the radial dependence of the torque that created the warp, and
thereby provide information on the shape of the halo at large radii.

Because the two separate components of the torque in our model cause
differential precessions in opposite senses, the net retrograde
precession of the warp is less differential than when only one is
present.  The moment we removed the external torque, the warp began to
wind, although its amplitude did not decrease.  Thus the quasi-steady
warp we observe that lasts for more than a Hubble time is a
consequence of steady external forcing.

Even though the disc precesses due to the external torque, its motion
is hardly damped over many Gyr, in contrast to the expectations
from \citet{nel_tre_95}.  Damping is weak because the slow precession
rate allows the inner halo to remain closely aligned with the disc,
which therefore causes little drag.  The very weak damping seems to be
caused more by the relative precession of the inner and outer halo.

A fixed outer torus is clearly unrealistic and external forcing is
likely to be strongly time-dependent in reality.  We cannot say much
about the rate at which misalignments in real halos will settle, but
we have shown that the warp survives for a few gigayears after the
torus is removed.  Time dependence may anyway be a side issue if the
halo axis is constantly slewing, as argued by \citet{qui_bin_92}.
Warps formed this way can be repeatedly regenerated when a new infall
event happens.  Since cosmic infall and mergers are more likely to
happen in a denser environment, warps can be induced more frequently
in such an environment, which is consistent with the warp statistics
in \citet{gar_etal_02}.

Since warps are ubiquitous and a gravitational phenomenon closely
related to hierarchical galaxy formation, their properties may
possibly be able to constrain small-scale cosmic structure.  Clearly
further experimentation with various external perturbations, halo
profiles and masses, disc properties (random motions, barredness),
etc., is required to discover what might be inferred from the
existence of warps in galaxies.

\section{Acknowledgments}
We thank Roelof Bottema for supplying the image used in Fig. 1.
Stimulating discussions with James Binney, Alar Toomre, Scott
Tremaine, and Victor Debattista have helped our understanding
considerably.  We also thank John Kormendy and an anonymous referee
for helpful suggestions.  This work was supported by NSF grants
AST-0098282 and AST-0507323 to JAS and a Bevier Fellowship from Rutgers
University to JS.


\appendix
\onecolumn

\section{The precession rate due to the torus}
\label{app:pre}

\begin{figure} 
\centerline{\includegraphics[angle=0, width=.5\hsize]{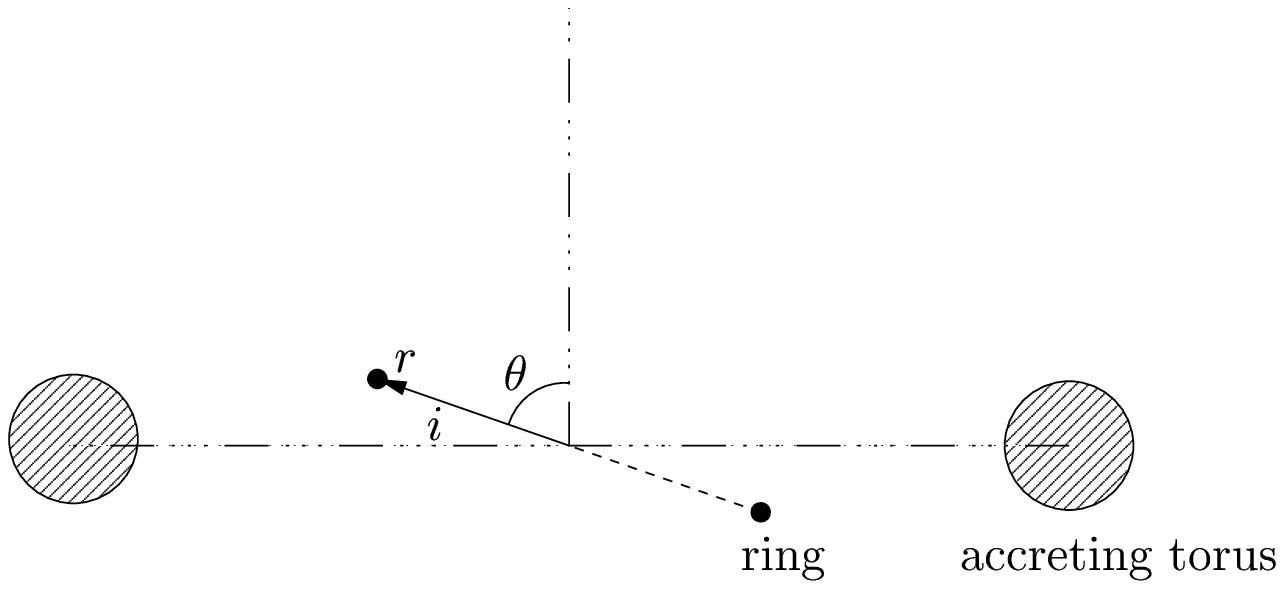}}
\caption{The calculation of the precession rate of a disc annulus, at
    radius $r$ and inclined relative to the torus symmetry plane with
    the angle $i$, under the torque of the accreting torus. The
    precession is retrograde about the symmetry axis of the torus.}
\label{fig:torquefig}
\end{figure}

Suppose the disc is composed of many rigid rings (or annuli). Here we
calculate the retrograde precession rate of one such ring, at radius
$r$ and inclined relative to the torus symmetry plane with the angle $i$, due to the potential of the accreting torus (see Figure~\ref{fig:torquefig}).

The potential of the accreting torus (its thickness ignored) can be obtained as the following, by solving Laplace's equation.
\begin{equation}
\label{eqn:toruspot}
\Phi(r, \; \theta) = -\GMoR \left[ 1 - \frac{1}{2} \left( \roRt \right)^2 \Ptwo
                               + \frac{3}{8} \left( \roRt \right)^4 P_4(\cos \theta)
			       + O\left( r^6/\Rt^6 \right)
			       \right],
\end{equation}
where $\theta$ is the conventional polar angle in the polar coordinate
system based on the torus; $\Rt$ and $\Mt$ are the radius and mass of the torus, respectively. 

We define
\begin{equation}
T(r) = \frac{1}{2} \GMoR \left( \roRt \right)^2 
\quad {\rm and} \quad
Q(r) = \frac{3}{8} \GMoR \left( \roRt \right)^4
\end{equation}
for convenience, so to the order of $r^4/\Rt^4$ we have
\begin{equation}
\Phi(r, \; \theta) = -\GMoR + T(r) \Ptwo - \; Q(r) \Pfour.
\end{equation}

The only force component relevant to the calculation of torque on the annulus is  $F_\theta$:
\begin{equation}
F_\theta  = - \frac{1}{r} \frac{\partial\Phi}{\partial\theta}
          = - \frac{T(r)}{r} \frac{d\Ptwo}{d\cos \theta} \cdot (-\sin \theta) + \frac{Q(r)}{r} \frac{d\Pfour}{d\cos \theta} \cdot (-\sin \theta).
\end{equation}

Since
\begin{eqnarray*}
P_2(x)=\frac{1}{2}(3x^2-1); \quad  \frac{dP_2(x)}{dx} = 3x; 
\end{eqnarray*}
\begin{eqnarray*}
P_4(x)=\frac{1}{8}(35x^4-30x^2+3); \quad  {\rm and} \quad 
\frac{dP_4(x)}{dx} = \frac{1}{2}(35x^3-15x),
\end{eqnarray*}

we find
\begin{equation}
F_\theta  = 3 \frac{T(r)}{r} \cos \theta \; \sin \theta + \frac{1}{2} \frac{Q(r)}{r} (15\cos\theta - 35\cos^3\theta) \sin \theta.
\end{equation}

The torque exerted on a point mass in the rigid ring due to the torus
depends on the azimuth of the point mass; the maximum value of the
torque occurs at azimuth $\phi = 0$ or $\pi$ where $\theta=\pi/2 - i$.
\begin{eqnarray}
& \tau_{\rm max} \!\!\!\! &= |{\mathbf r } \times {\mathbf F }| = r F_\theta
= 3 T(r) \cos \theta \; \sin \theta + \frac{1}{2}  Q(r) (15 - 35\cos^2\theta) \cos\theta \; \sin\theta \nonumber \\
& &= 3 T(r) \sin i \; \cos i + \frac{1}{2}  Q(r) (15 - 35\sin^2 i) \sin i \; \cos i.
\end{eqnarray}

The azimuthally average torque on the rigid ring is therefore
\begin{equation}
\tauave = \frac{\int_0^{2\pi} \tau_{\rm max} \cos^2 \phi \;
  d\phi}{2\pi} = \frac{1}{2} \tau_{\rm max} = \frac{3}{2}T(r) \sin i
  \; \cos i + \frac{1}{4} Q(r) (15 - 35\sin^2 i) \sin i \; \cos i.
\label{eqn:torque_2nd}
\end{equation}

We can obtain the total retrograde precession rate due to the torus
(up to the fourth order):
\begin{eqnarray}
&\omegap &= \frac{\tauave}{L\sin i} = \frac{3}{2} \frac{T(r)}{r \cdot \Vc} \cos i + \frac{1}{4} \frac{Q(r) (15 - 35\sin^2 i)}{r \cdot \Vc} \cos i \nonumber \\
&        &= \frac{3}{4} \frac{G\Mt r}{\Rt^3 \Vc} \cos i \left[ 1+ \frac{1}{8} \left( \roRt \right)^2 (15 - 35\sin^2 i) \right] \nonumber \\
&        &=\omegap^{(2)} \left[ 1+ \frac{1}{8} \left( \roRt \right)^2 (15 - 35\sin^2 i) \right]. \label{eqn:omegap_total}
 \end{eqnarray}

For qualitative purposes, we generally use the following simpler second-order approximation ($\omegap^{(2)}$), unless the radius of the annulus is large.
\begin{equation}
\label{eqn:omegap_2nd}
\omegap \approx \omegap^{(2)} = \frac{3}{4} \frac{G\Mt r}{\Rt^3 \Vc} \cos i.
\end{equation}

Note that same results can be obtained for the nodal precession
rate of a test particle's circular orbit (as opposed to a rigid ring),
the only difference here is that the average is taken as the time
average over one orbit \citep[the orbit-average procedure can be found
in][Appendix]{steima_84}.

\section{The derivation of tip angle of rings and azimuthal angle of their LON}
\label{app:thetaphi}

\begin{figure} 
\centerline{\includegraphics[angle=0, width=.7\hsize]{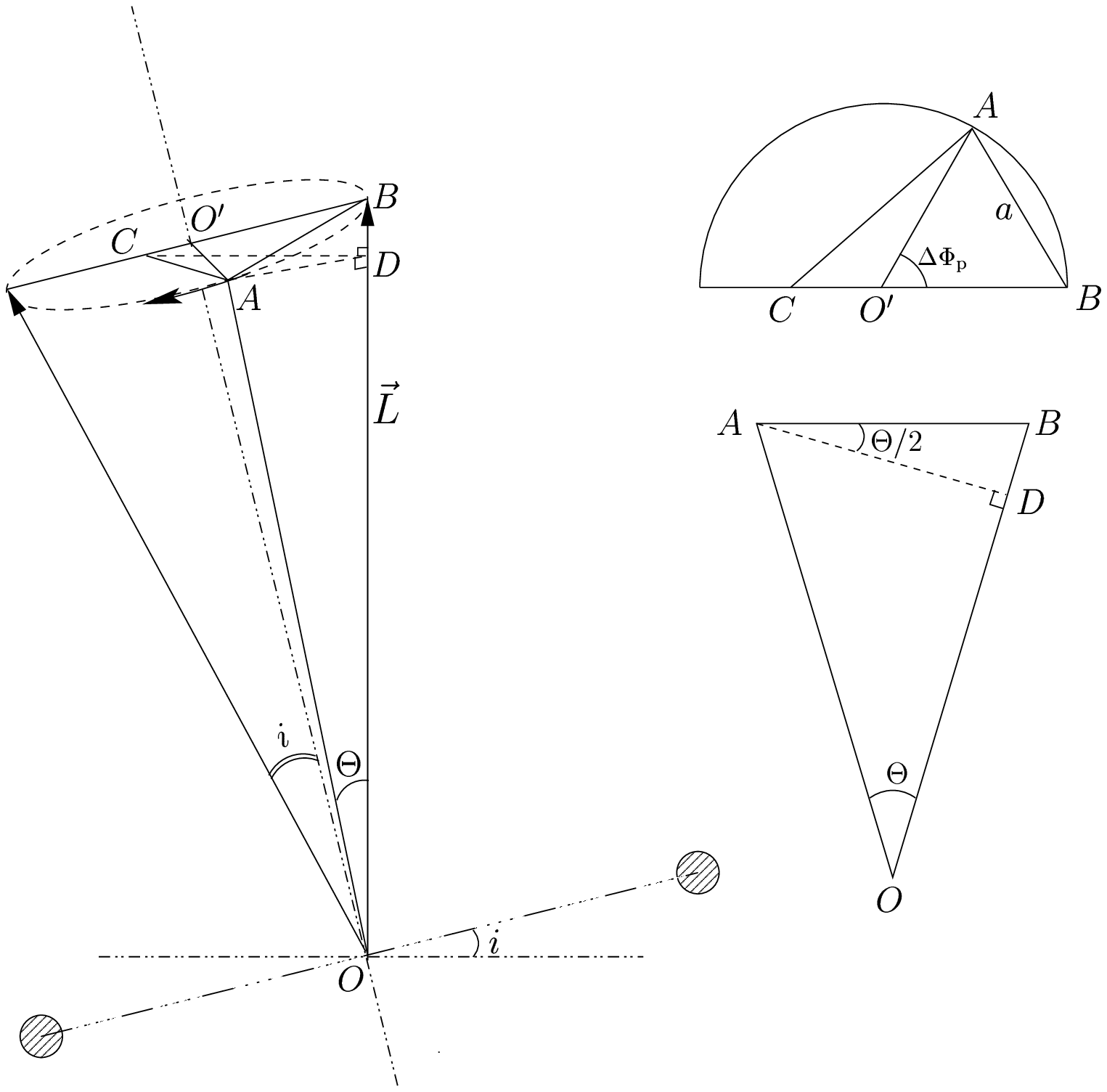}}
\caption{The schematic diagram showing how $\Theta$ and $\Philon$ 
    are calculated. }
\label{fig:thetaphi_ang}
\end{figure}

We first calculate the tip angle $\Theta$ relative to the original $z$-axis. We know $\AB = 2\overline{OB} \sinht$. On the other hand we also have
\begin{equation}
\AB=2\overline{O^{\prime}B} \sinhp=2 \overline{OB} \sini \sinhp 
\end{equation}
here $\dphi= \int \omegap(t) \; dt$ is angle that the annulus has precessed. So obviously the equation to determine $\Theta$ is
\begin{equation}
\label{eqn:sinht}
\sinht = \sini \sinhp
\end{equation}

If we define the azimuthal angle of the LON of a ring as the counter-clockwise angle from $+x$ axis, $\Phi_{\rm LON} = \angle ADC - \pi$. It can be shown, with some trigonometry on triangles in Figure~\ref{fig:thetaphi_ang}, that 
\begin{equation}
\angle ADC = \cos^{-1} \left[ \frac {\cosi \sinhp} {\sqrt{1 - \sinis \sinhps} }\right].
\end{equation}

So
\begin{equation}
\Phi_{\rm LON} = \angle ADC - \pi = \cos^{-1} \left[ \frac {\cosi \sinhp} {\sqrt{1 - \sinis \sinhps}}\right] -\pi \\
\label{eqn:philon}
\end{equation}

For small $i$, we have
\begin{equation}
\Phi_{\rm LON} \approx -\frac{\pi}{2} - \dphih 
\label{eqn:philon_app}
\end{equation}

\bsp

\label{lastpage}

\end{document}